\pdfoutput=1

\documentclass[acmlarge]{acmart}
\usepackage{subfigure}
\usepackage{booktabs} 
\usepackage{makecell}
\usepackage{multirow}
\usepackage{xfrac}
\usepackage{graphicx}
\usepackage{color}
\newcommand{\blue}{}

\usepackage[ruled]{algorithm2e} 

\SetAlFnt{\small}
\SetAlCapFnt{\small}
\SetAlCapNameFnt{\small}
\SetAlCapHSkip{0pt}
\IncMargin{-\parindent}

\acmJournal{IMWUT}
\acmVolume{2}
\acmNumber{3}
\acmArticle{135}
\acmYear{2018}
\acmMonth{9}
\acmArticleSeq{11}

\setcopyright{acmcopyright}

\acmPrice{15.00}
\acmDOI{10.1145/3264945}

\received{February 2018}
\received{May 2018}
\received[accepted]{September 2018}

\begin{document}
\title{Deep Room Recognition Using Inaudible Echos}

\author{Qun Song}
\affiliation{Energy Research Institute, Interdisciplinary Graduate School, Nanyang Technological University, Singapore}
\affiliation{School of Computer Science and Engineering, Nanyang Technological University, Singapore}
\author{Chaojie Gu}
\affiliation{School of Computer Science and Engineering, Nanyang Technological University, Singapore}
\author{Rui Tan}
\affiliation{School of Computer Science and Engineering, Nanyang Technological University, Singapore}

\authorsaddresses{%
  Authors' addresses: Qun Song, song0167@e.ntu.edu.sg; Chaojie Gu, gucj@ntu.edu.sg; Rui Tan, tanrui@ntu.edu.sg, School of Computer Science and Engineering, Nanyang Technological University, Block N4 \#02a-32, 50 Nanyang Avenue, Singapore 639798.
}

\begin{abstract}
  Recent years have seen the increasing need of location awareness by mobile applications. This paper presents a room-level indoor localization approach based on the measured room's echos in response to a two-millisecond single-tone inaudible chirp emitted by a smartphone's loudspeaker. Different from other acoustics-based room recognition systems that record full-spectrum audio for up to ten seconds, our approach records audio in a narrow inaudible band for 0.1 seconds only to preserve the user's privacy. However, the short-time and narrowband audio signal carries limited information about the room's characteristics, presenting challenges to accurate room recognition. This paper applies deep learning to effectively capture the subtle fingerprints in the rooms' acoustic responses. Our extensive experiments show that a two-layer convolutional neural network fed with the spectrogram of the inaudible echos achieve the best performance, compared with alternative designs using other raw data formats and deep models. Based on this result, we design a {\em RoomRecognize} cloud service and its mobile client library that enable the mobile application developers to readily implement the room recognition functionality without resorting to any existing infrastructures and add-on hardware.
  Extensive evaluation shows that RoomRecognize achieves 99.7\%, 97.7\%, 99\%, and 89\% accuracy in differentiating 22 and 50 residential/office rooms, 19 spots in a quiet museum, and 15 spots in a crowded museum, respectively. Compared with the state-of-the-art approaches based on support vector machine, RoomRecognize significantly improves the Pareto frontier of recognition accuracy versus robustness against interfering sounds (e.g., ambient music).
\end{abstract}

%
%

%
%


\maketitle

\renewcommand{\shortauthors}{Q. Song et al.}

\section{Introduction}
\label{sec:intro}

Recent years have seen the increasing need of location awareness by mobile applications. As of November 2017, 62\% of the top 100 free Android Apps on Google Play require location services. While GPS can provide outdoor locations with satisfactory accuracy, determining indoor locations has been a hard problem. Research in the last decade has proposed a plethora of indoor localization approaches that use various signals such as Wi-Fi \cite{bahl2000radar,haeberlen2004practical}, GSM \cite{hightower2005learning}, FM radio \cite{chen2012fm}, geomagnetism \cite{chung2011indoor}, and aircraft ADS-B messages \cite{eichelberger2017}. These systems aim at achieving meters to centimeters localization accuracy. Differently, this paper aims to design a practical room-level localization approach for off-the-shelf smartphones using their built-in audio systems only.
Room-level localization is desirable in a range of ubiquitous computing applications. For instance, in a hospital,
knowing which room that a patient is in is important to responsive medical aid when the patient develops an emergent condition (e.g., falling in a faint). In a museum, knowing which exhibition chamber that a tourist is in can largely assist the automation of her multimedia guide that is often provided as a mobile App nowadays.
In a smart building, the room-level localization of the residents can assist the automation of illumination and air conditioning to improve energy efficiency and occupant comfort.

The requirements of existing indoor localization approaches can be summarized as: ({\bf R1}) a dedicated or an existing infrastructure that provides signals for localization \cite{haeberlen2004practical,hightower2005learning,chen2012fm}; ({\bf R2}) add-on equipment to the user's smartphone \cite{eichelberger2017}; and/or ({\bf R3}) a training process to collect data for the subsequent localization processes \cite{haeberlen2004practical,chung2011indoor}. Any of the above three requirements leads to a certain degree of overhead in deploying the indoor localization services. However, most existing approaches have at least one of the above drawbacks.
In respect of the requirements {\bf R1} and {\bf R2}, as our approach is based on the phone's built-in audio only, it does not require any infrastructure and add-on equipment to the phone. If effective acoustic representations of the target rooms can be found, acoustics-based room-level localization can be cast into a supervised multiclass classification problem by treating the rooms as classes. Thus, in respect of the last requirement {\bf R3}, we aim to design an acoustics-based room recognition approach with an easy training data collection process.
For instance, the system trainer can simply carry a smartphone to the target rooms, click some buttons on the phone screen, and key in the room names.
Such a process can be easily accomplished by non-experts. Thus, compared with other fingerprint-based approaches \cite{haeberlen2004practical,chung2011indoor} that require precisely controlled training data collection processes at dense locations,
the training of our system is practical and nearly effortless. Once trained, for the end users, the room recognition becomes an out-of-the-box feature.



Some existing indoor localization systems have incorporated acoustic sensing. An early study, SurroundSense \cite{azizyan2009surroundsense}, used the acoustic loudness in combination of other sensing modalities such as imaging to distinguish the ambient environments. However, the trials of using acoustics alone have not achieved satisfactory performance yet. For instance, with acoustics only, SurroundSense outperforms barely random guessing \cite{tarzia2011indoor}. Batphone \cite{tarzia2011indoor} used the acoustic background spectrum as a feature to classify rooms. However, it achieves a 69\% accuracy only in classifying 33 rooms. Moreover, as it uses the $[0, 7\,\text{kHz}]$ audible band, it is inevitably susceptible to foreground sounds. As shown in \cite{tarzia2011indoor} and this paper, Batphone fails in the presence of chatters and ambient music. Different from these systems \cite{azizyan2009surroundsense,tarzia2011indoor} that passively listen to the room's foreground or background sounds, in this paper, we investigate an active sensing scheme that uses the phone's loudspeaker to emit a predefined signal and then uses the microphone to capture the reverberation in the measured room. Intuitively, due to different sizes of the rooms and different acoustic absorption rates of the wall and furniture materials, the acoustic echos may carry features for distinguishing rooms. Moreover, as we use an {\em a priori} signal to stimulate the room, we can design the signal to minimize the unwanted impacts of other interfering sounds (e.g., ambient music and human conversations) on the room recognition.

However, the following two basic requirements present challenges to the design of the active room recognition system. First, lengthy audio recording in private/semi-private spaces (e.g., homes and wards) to capture acoustic features may cause the user's privacy concern. Therefore, to avoid privacy breach, the audio recording time needs to be minimal. Second, it is desirable to use inaudible sounds with frequencies above $20\,\text{kHz}$ as the stimulating signals. This avoids annoyance to the user and well separates the stimulating signals from most man-made audible sounds to improve the system's robustness.
Moreover, as the performance (e.g., sensitivity) of most smartphone audio systems decreases drastically with the frequency beyond $20\,\text{kHz}$, it is desirable to use a narrowband stimulating signal with a central frequency close to $20\,\text{kHz}$. Therefore, to meet the above two requirements, we should use a short-time narrowband inaudible stimulating signal.
However, due to the limited time and frequency spans of the stimulating signal, the responding echos will inevitably carry limited information about the measured room.
As a result, extracting features from the echos, which is a key step of supervised classification, to distinguish the rooms becomes challenging. In particular, classification system designers generally handcraft the features, often through exhaustive trials of popular features.
This {\em ad hoc} approach, however, is ineffective if the distinguishability is intricately embedded in the raw data.


The emerging {\em deep learning} method \cite{lecun2015deep} automates the design of feature extraction by unsupervised feature learning and employs deep models with a large number of parameters to effectively capture the intricate distinguishability in the raw data. Its outperforming performance has been demonstrated in a number of application domains such as image classification \cite{krizhevsky2012imagenet}, speech recognition \cite{hinton2012deep}, natural language understanding \cite{collobert2011natural}, and etc.
Thus, deep learning is a promising method to address the aforementioned challenges caused by the need of using short-time narrowband inaudible stimulating signals in the active sensing scheme. This paper presents the design of a deep room recognition approach through 
extensive experiments investigating appropriate forms of the raw data, the choice of the deep model, and the design of the model's hyperparameters. The results show that a two-layer convolutional neural network (CNN) fed with the spectrogram of the captured inaudible echos achieves the best performance. In particular, based on a $100\,\text{ms}$ audio recording after a $2\,\text{ms}$ $20\,\text{kHz}$ single-tone chirp, the CNN gives 99.7\%, 99\%, and 89\% accuracy in distinguishing 22 residential and office rooms, 19 spots in a quiet museum, and 15 spots in a crowded museum, respectively.
Moreover, it scales well with the number of rooms -- it maintains a 97.7\% accuracy in distinguishing 50 rooms.
Our approach significantly outperforms the passive-sensing-based Batphone \cite{tarzia2011indoor} that achieves a 69\% accuracy using ten seconds of privacy-breaching audio recording. Moreover, compared with a state-of-the-art active sensing approach based on support vector machine (SVM) \cite{rossi2013roomsense}, our CNN-based approach significantly improves the Pareto frontier of recognition accuracy versus robustness against interfering sounds (e.g., ambient music).
Based on these results, we design a cloud service named {\em RoomRecognize} that facilitates the integration of our room recognition service into mobile applications. In particular, RoomRecognize supports a participatory learning mode where the end users can contribute training data.

The contributions of this paper include (i) an in-depth measurement study on the rooms' acoustic responses to a short-time single-tone inaudible chirp, (ii) the design of a deep model that effectively captures the subtle differences in rooms' acoustic responses, (iii) extensive evaluation of our approach in real-world environments including homes, offices, class rooms, and museums, as well as (iv) a room recognition cloud service and its mobile client library that are ready for application integration.

The rest of this paper is organized as follows. Section~\ref{sec:related} reviews related work. Section~\ref{sec:measurement} presents a measurement study to understand rooms' responses to inaudible chirps. Section~\ref{sec:approach} presents the design of our deep room recognition system. Sections~\ref{sec:rr} and \ref{sec:eval} design and evaluate RoomRecognize, respectively.
Section~\ref{sec:discuss} discusses several issues not addressed in this paper.
Section~\ref{sec:conclude} concludes.

\section{Related Work}
\label{sec:related}

As a challenging problem, indoor localization has received extensive research.
Existing approaches are either {\em infrastructure-dependent} or {\em infrastructure-free}.
The infrastructure-dependent approaches leverage existing or pre-deployed infrastructures to determine the location of a mobile device. Existing radio frequency (RF) infrastructures such as 802.11 \cite{bahl2000radar,haeberlen2004practical}, cellular \cite{hightower2005learning}, FM radios \cite{chen2012fm}, aircraft automatic dependent surveillance-broadcast (ADS-B) systems \cite{eichelberger2017}, and a combination of multiple schemes \cite{duuniloc} have been used for indoor localization. The 802.11-based approaches require dense deployment of access points (APs). Most RF-based approaches are susceptible to the inevitable fluctuations of the received signal strength due to complex signal propagation and/or adaptive transmit power controls \cite{haeberlen2004practical,chen2012fm}. The reception of ADS-B signals needs special hardware that is not available on commodity mobile devices. Existing studies have also proposed to use acoustic infrastructure for indoor localization. WALRUS \cite{borriello2005walrus} uses desktop computers to emit inaudible acoustic beacons to localize mobile devices.
Scott and Dragovic \cite{scott2005audio} use multiple microphones deployed in a space to determine the locations of human sounds such as clicking fingers. However, the dedicated, laborious deployments of the acoustic infrastructures impede the adoption of these approaches.

Infrastructure-free approaches leverage location indicative signals including geomagnetism \cite{chung2011indoor}, imaging \cite{dong2015imoon}, acoustics, and etc. As our work uses acoustics, the following literature survey focuses on the acoustics-based approaches that can be classified into {\em passive} and {\em active} sensing approaches.

The passive acoustic sensing analyzes the ambient sounds only to estimate the mobile device's location. SurroundSense \cite{azizyan2009surroundsense} uses multiple phone's sensors (microphone, camera, and accelerometer) to distinguish the ambient environments (e.g., different stores in a mall). It uses the loudness as the main acoustic feature. As tested in \cite{tarzia2011indoor}, with the acoustic modality only, SurroundSense outperforms barely random guessing. Its lengthy acoustic recording and image capture also raises privacy concerns when used in private spaces. Batphone \cite{tarzia2011indoor} applies the nearest neighbor algorithm to recognize a room based on the acoustic background spectrum in the $[0, 7\,\text{kHz}]$ band. In quiet environments, Batphone achieves a 69\% accuracy in classifying 33 rooms. However, it is highly susceptible to foreground sounds. Using the $[0, 7\,\text{kHz}]$ band, it performs worse than random guessing in the presence of a single human speaker. Narrowing the band to $[0, 300\,\text{Hz}]$ rescues Batphone to achieve a 63.4\% accuracy, but drags the quiet case's accuracy to 41\% \cite{tarzia2011indoor}. Thus, Batphone has a poor Pareto frontier of recognition accuracy versus robustness against interfering sounds.

The active acoustic sensing uses the phone's loudspeaker to emit chirps and the microphone to capture the echos of the measured space. This approach has been applied to semantic location recognition \cite{kunze2007symbolic,fan2014public,tachikawa2016predicting}. In \cite{kunze2007symbolic}, a decision tree is trained to classify a phone's semantic location (e.g., in a backpack, on a desk, in a drawer, etc) using active vibrational and acoustic sensing. The phone emits eight audible multi-tone chirps that cover a frequency range from $0.5\,\text{kHz}$ to $4\,\text{kHz}$.
In \cite{fan2014public}, the mel-frequency cepstral coefficients (MFCC) \cite{zheng2001comparison} of the acoustic echos triggered by audible sine sweep chirps are used to detect whether the phone's environment is a restroom.
In \cite{tachikawa2016predicting}, active acoustic sensing, combined with other passive sensing using magnetometer and barometer, classifies the phone's environment into six semantic locations: desk, restroom, meeting room, elevator, smoking area, and cafeteria. The classification is based on a decision tree trained by the random forest algorithm with MFCC of the audible echos as the acoustic feature. These semantic localization approaches are fundamentally different from our room recognition approach, in that they give the type of the context only and they do not tell the room's identity. For instance, the approaches in \cite{fan2014public,tachikawa2016predicting} do not differentiate the restrooms in different buildings. Following the active acoustic sensing approach, EchoTag \cite{tung2015echotag} determines the phone's position using SVM among a set of predefined positions that are fingerprinted using audible echos.
In other words, EchoTag ``remembers'' predefined positions with certain tolerance ($0.4\,\text{cm}$) and resolution ($1\,\text{cm}$). This is different from our objective of room recognition.

RoomSense \cite{rossi2013roomsense} is a system that is the closest to ours.
Following the active sensing approach, a RoomSense phone emits an audible sound of 0.68 seconds and classifies a room using SVM based on the echos' MFCC features.
As RoomSense uses the whole audible band, it is susceptible to ambient sounds. Thus, it demands well controlled conditions, e.g., closed windows and doors \cite{rossi2013roomsense}. In contrast, due to the use of narrowband stimulating signal, our approach is much more robust against ambient sounds. In this paper, we conduct experiments to extensively compare our approach with RoomSense and our improved versions of RoomSense that use narrowband stimulating signals as well. The results show that, when the acoustic sensing is restricted to a narrow inaudible band, our spectrogram-based CNN gives 22\% and 17.5\% higher recognition accuracy than RoomSense's MFCC-based SVM, in the absence and presence of interfering ambient music, respectively.


Active acoustic sensing has also been used for ranging, moving object tracking, and gesture recognition. BeepBeep \cite{peng2007beepbeep} and SwordFight \cite{zhang2012swordfight} measure the distance between two phones by acoustic ranging.
Recent studies also apply active acoustic sensing to track the movements of a finger \cite{wang2016device}, breath \cite{nandakumar2015contactless}, and a human body using inaudible chirps embedded in music \cite{CovertBand}. However, these studies \cite{peng2007beepbeep,zhang2012swordfight,wang2016device,nandakumar2015contactless,CovertBand} address ranging and ranging-based moving object tracking, rather than classification. SoundWave \cite{gupta2012soundwave} generates inaudible tones with a commodity device's built-in speaker and analyzes the Doppler-shifted reflections sensed by a built-in microphone to infer various features such as velocity, orientation, proximity, and size of moving hands. Based on these features, SoundWave recognizes the hand gesture.

\section{Measurement Study}
\label{sec:measurement}

\begin{figure}
  \centering
  \begin{minipage}[t]{.55\textwidth}
    \includegraphics[width=\textwidth]{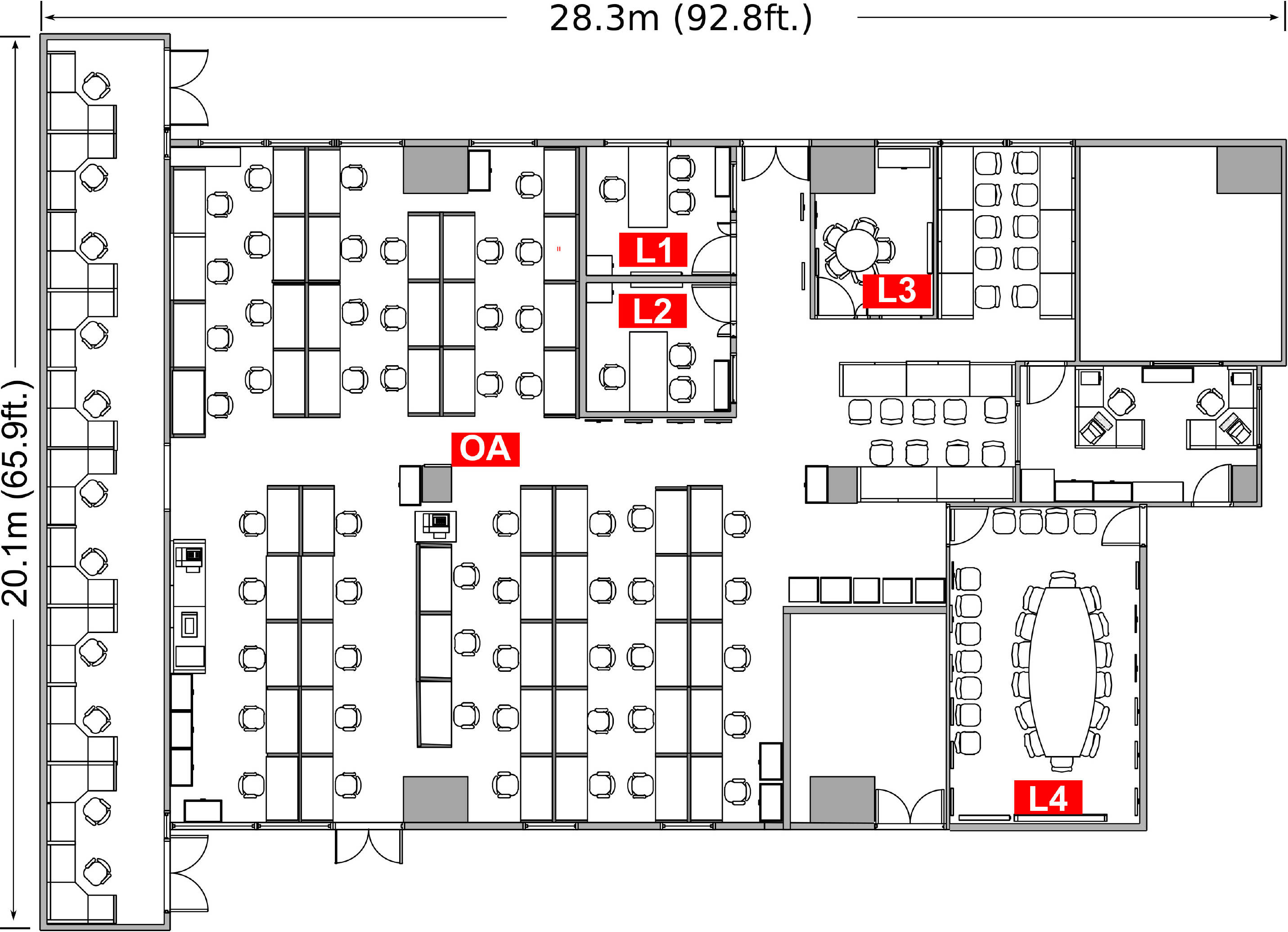}
    \caption{Floor plan of the lab.}
    \label{fig:lab-floor-plan}
  \end{minipage}
  \hspace{0.01\textwidth}
  \begin{minipage}[t]{.43\textwidth}
    \includegraphics{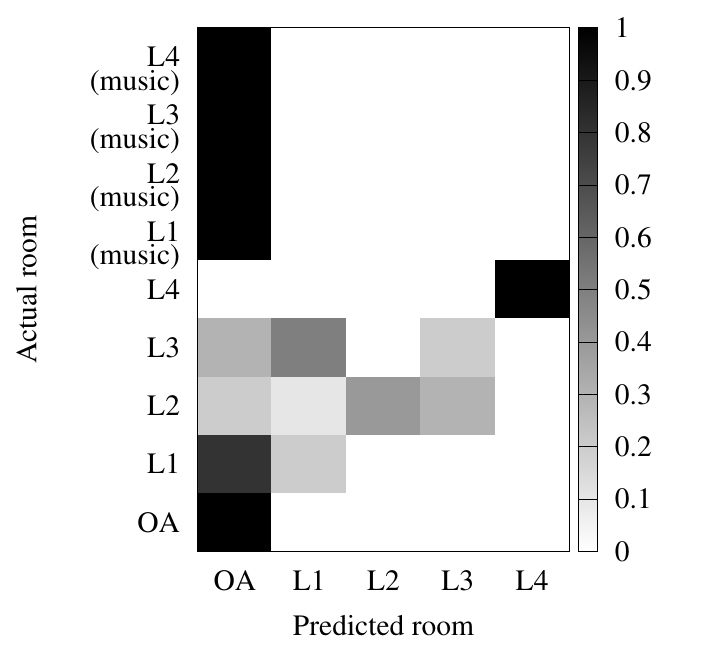}
    \caption{\blue Confusion matrix of Batphone \cite{tarzia2011indoor} in the lab.}
    \label{fig:batphone-confusion}
  \end{minipage}
\end{figure}

In this section, we conduct measurements to motivate our study and gain insights for the design of our approach. The measurements are conducted in a computer science lab shown in Fig.~\ref{fig:lab-floor-plan}. The measured rooms are labeled by ``Lx''. The open area of the lab is labeled by ``OA''.

\subsection{Performance of Passive Acoustic Sensing}
\label{subsec:batphone-test}

As discussed in Section~\ref{sec:related}, Batphone is a recent room recognition approach based on passive acoustic sensing. We install the implementation of Batphone \cite{tarzia2011indoor} from Apple's App Store \cite{batphone-impl} on an iPhone 6s. We test its performance using five rooms, i.e., L1 to L4, and OA shown in Fig.~\ref{fig:lab-floor-plan}. We use the default setting of Batphone to collect training data in each room. Specifically, the training data collection in each room takes ten seconds. During the testing phase, we test Batphone for ten times in each room, in the morning, afternoon, and evening. Thus, Batphone is tested for each room for 30 times totally. Note that the data collection for each test takes ten seconds. As discussed in \cite{tarzia2011indoor}, Batphone has significant performance drop in the presence of foreground sounds. Thus, in the first set of tests, we keep quiet environment in favor of Batphone during the training and testing phases. The bottom part of Fig.~\ref{fig:batphone-confusion} shows Batphone's confusion matrix in the first set of tests. When the actual room is OA and L4, Batphone can accurately recognize the two rooms. However, when the actual room is L1, L2, or L3, Batphone yields high recognition errors. For example, when the actual room is L2, Batphone gives a 40\% accuracy only. A possible reason for such low performance is that, as these rooms are in proximity with each other, they may have similar ambient background spectrum that Batphone relies on. In the second set of tests, we evaluate the performance of Batphone in the presence of foreground sounds. Specifically, we keep quiet environment during the training phase and play a music sound track on a laptop computer during the testing phase. The top part of Fig.~\ref{fig:batphone-confusion} shows Batphone's confusion matrix in this set of tests. The rooms L1 to L4 are always wrongly classified as OA. The above two sets of tests show the challenges faced by passive acoustic sensing in real-world environment and its susceptibility to interfering sounds.

\subsection{Rooms' Responses to Single-Tone Chirps}
\label{sec:responses}


\begin{figure}
  \centering
  \includegraphics[width=.5\textwidth]{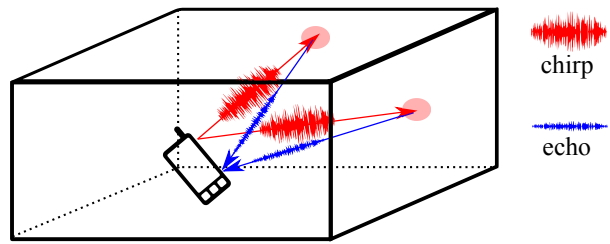}
  \caption{Active acoustic sensing.}
  \label{fig:room-model}
\end{figure}

The results in Section~\ref{subsec:batphone-test} motivate us to explore active acoustic sensing. Fig.~\ref{fig:room-model} illustrates the active sensing scheme. Specifically, the smartphone uses its loudspeaker to emit an acoustic chirp and meanwhile uses its microphone to capture the measured room's response. In this section, we conduct a small-scale measurement study to obtain insightful observations on the rooms' responses. These observations help us make various design choices for an effective active sensing approach in Section~\ref{sec:approach}. Note that the systematic evaluation on the effectiveness of our active sensing approach will be presented in Section~\ref{sec:eval}.


\subsubsection{Measurement Setup}
\label{subsubsec:measurement-setup}

Our measurement study uses a Samsung Galaxy S7 phone with Android 7.0 Nougat. To collect data, we develop a program that emits a chirp with a time duration of $2\,\text{ms}$ using the phone's loudspeaker every $100\,\text{ms}$. Meanwhile, the program continuously samples the phone's microphone at a rate of $44.1\,\text{ksps}$ and stores the raw data to the phone's internal memory for offline analysis. Thus, the program will capture both the chirps that directly propagate from the loudspeaker and the echos from the room if any. With the setting of $2\,\text{ms}$ for the chirp length, the chirp will not overlap the echos from the objects that are more than $34\,\text{cm}$ from the phone.
By emitting the chirp every $100\,\text{ms}$, in each measured room, we can easily collect a large volume of the room's acoustic responses to the chirps to drive the design of our deep learning based approach. We set the period to be $100\,\text{ms}$, because from our preliminary measurements, the echos vanish after $100\,\text{ms}$ from the chirp. In each room, we randomly select at least two spots. We place the phone at each spot and run the program to collect data for about half an hour. Note that, this half an hour time period is merely for collecting data to understand the rooms' responses with sufficient statistical significance. The minimum needed volume of training data for our room recognition system will be investigated in Section~\ref{sec:eval}.


\begin{figure}
  \centering
  \begin{minipage}[t]{.75\textwidth}
    \vspace{-1.7in}
    \includegraphics{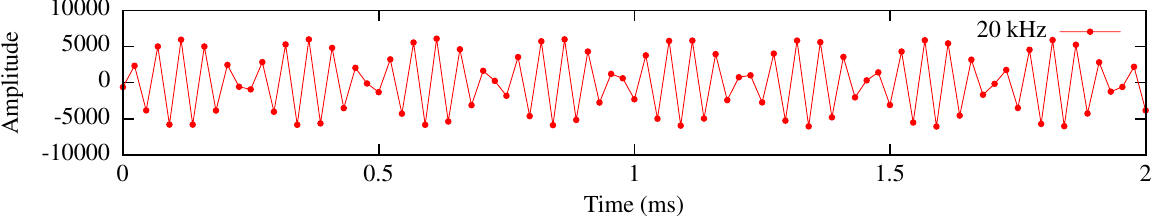}
    \includegraphics{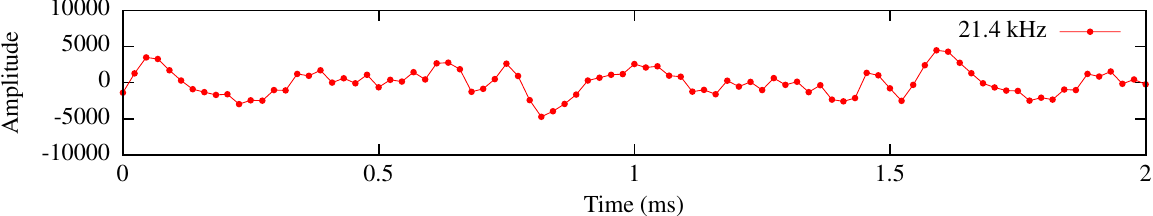}
    \caption{$20\,\text{kHz}$ and $21.4\,\text{kHz}$ chirps received by the phone's microphone.}
    \label{fig:inaudibles}
  \end{minipage}
  \begin{minipage}[t]{.24\textwidth}
    \includegraphics{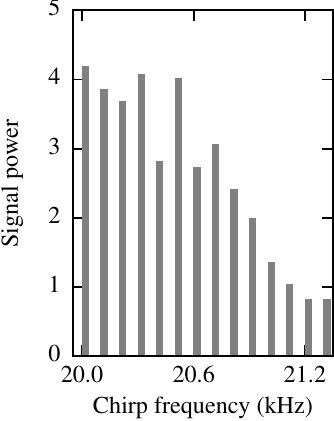}
    \caption{Chirp power.}
    \label{fig:inaudibles-power}
  \end{minipage}
\end{figure}

Existing studies on active acoustic sensing often use sine sweep chirp \cite{fan2014public}, Maximum Length Sequence \cite{rossi2013roomsense}, and multi-tone chirp \cite{kunze2007symbolic} that cover a wide acoustic spectrum, including the audible range, to increase the information carried by the echos about the measured rooms. However, the audible chirps are annoying. In this paper, we propose to use a single-tone inaudible chirp to avoid the annoyance to the user and also improve the robustness of the room recognition system against interfering sounds. From our tests, the performance of the phone's audio system decreases with the frequency beyond $20\,\text{kHz}$. Fig.~\ref{fig:inaudibles} shows the signal recorded by the phone's microphone when the program emits $20\,\text{kHz}$ and $21.4\,\text{kHz}$ chirps. When the configured frequency is $20\,\text{kHz}$, the received signal does exhibit a $20\,\text{kHz}$ frequency. However, when the configured frequency is $21.4\,\text{kHz}$, the received signal is significantly distorted and becomes audible. This is because that the mechanical dynamics of either the loudspeaker or the microphone cannot well support such a high frequency. Fig.~\ref{fig:inaudibles-power} shows the power of the received signal versus the configured frequency. The decreasing trend indicates that the audio system's performance decreases with the frequency. Therefore, in this paper, we choose $20\,\text{kHz}$, i.e., the lowest inaudible frequency, for the stimulating signal used by our system. To check if a smartphone can emit and receive inaudible signals (e.g., the $20\,\text{kHz}$ tone used by our approach), utilities such as the Near Ultrasound Tests \cite{ultrasound-tests} provided by the Android Open Source Project and various tone generator and spectrum analyzer Apps in Apple's App Store can be used. From our tests, recent models of Android phones (e.g., Samsung Galaxy S7, S8, etc) and iPhone (6s, 7, and X) can well emit and receive the $20\,\text{kHz}$ tone used by our approach.


\subsubsection{Time-Domain Analysis}
\label{subsubsec:time-domain}

\begin{figure}
  \centering
  \subfigure[Acoustic trace in L3 for $100\,\text{ms}$ after the beginning of the chirp.]
  {
    \includegraphics{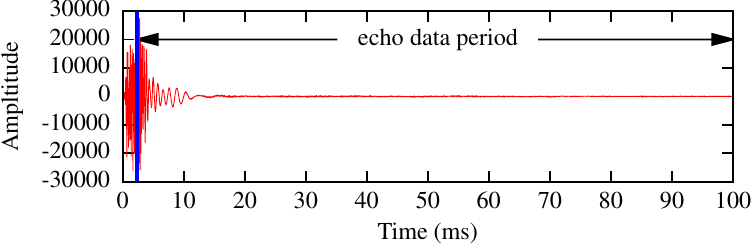}
    \label{fig:raw-l3}
  }
  \subfigure[Zoom-in view of the signal in (a) during (13.8, 100) ms.]
  {
    \includegraphics{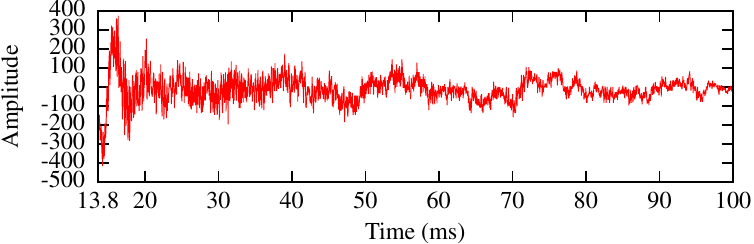}
    \label{fig:echo-l3}
  }
  \subfigure[Correlation with the chirp template in L3.]
  {
    \includegraphics{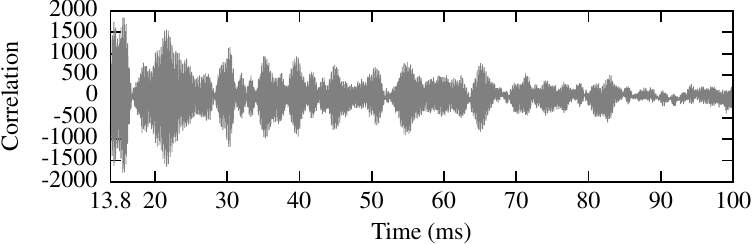}
    \label{fig:corr}
  }
  \subfigure[Correlation with the chirp template when tested outdoor.]
  {
    \includegraphics{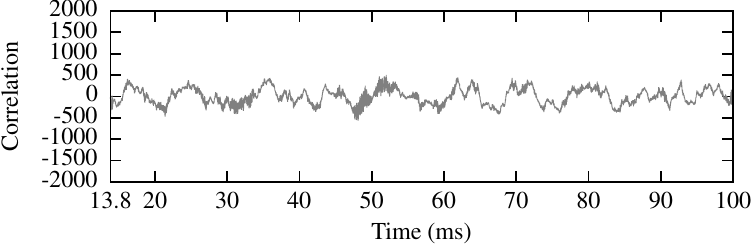}
    \label{fig:corr-outdoor}
  }
  \caption{Time-domain responses of room L3 and outdoor.}
\end{figure}

We analyze the data collected for a chirp in room L3 shown in Fig.~\ref{fig:lab-floor-plan}. The raw acoustic trace for $100\,\text{ms}$ is shown in Fig.~\ref{fig:raw-l3}. The period from the beginning to the first vertical line is the {\em chirp period} of $2\,\text{ms}$. During this period, the acoustic signal propagates directly from the phone's loudspeaker to its microphone. After the chirp period, we discard the data during $0.5\,\text{ms}$ as a safeguard region and use the data during the remaining $97.5\,\text{ms}$ to extract echos. This $97.5\,\text{ms}$ period is called {\em echo data period}. From Fig.~\ref{fig:lab-floor-plan}, we can see that the chirp does not immediately stop after the $2\,\text{ms}$ chirp period. It lasts for several milliseconds with decreasing amplitude.
Such damped oscillation can be caused by the mechanical dynamics of the loudspeaker's and the microphone's diaphragms.
As the damped oscillation still has much stronger intensity than that in the following time period that contains echos, to facilitate data visualization in this section, we discard the acoustic data collected within the first $13.8\,\text{ms}$ from the beginning of the chirp and use the data in the remaining period of $86.2\,\text{ms}$ to investigate the time-domain response of the measured room.
Fig.~\ref{fig:echo-l3} shows the zoom-in view of the signal in the echo data period.
By comparing Fig.~\ref{fig:raw-l3} and Fig.~\ref{fig:echo-l3}, we can see that the signal in the echo data period is about 100 times weaker, in terms of amplitude, than the self-heard chirp. The signal attenuation during propagation and the absorption by the surrounding objects are the main causes of the weak signals. Thus, we question the presence and salience of the echos in the signal shown in Fig.~\ref{fig:echo-l3}.

We slide a window of $2\,\text{ms}$ over the echo data period and compute the correlation between the sampled signal in each window and an ideal $20\,\text{kHz}$ sine wave template. Fig.~\ref{fig:corr} shows the correlation over time. We can clearly see wave packets, which indicate the presence of echos. In particular, there are more than ten interleaving strong and weak wave packets, which suggests multiple acoustic bouncebacks in the room. This shows that the phone's audio system can capture such intricate processes well. Fig.~\ref{fig:corr-outdoor} plots the correlation obtained outdoor. It does not show any wave packets, since there are no echos.


\subsubsection{Frequency-Domain Analysis}
\label{subsubsec:frequency-domain}

\begin{figure}
  \subfigure[Same room, same spot, different times]
  {
    \includegraphics{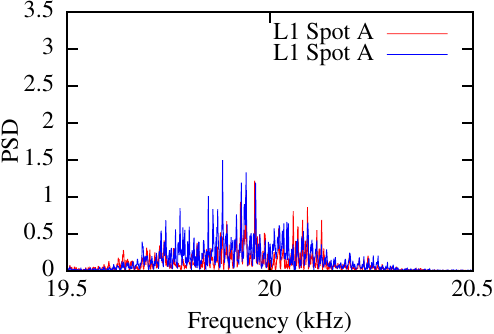}
    \label{fig:l1s6}
  }
  \subfigure[Same room, different spots]
  {
    \includegraphics{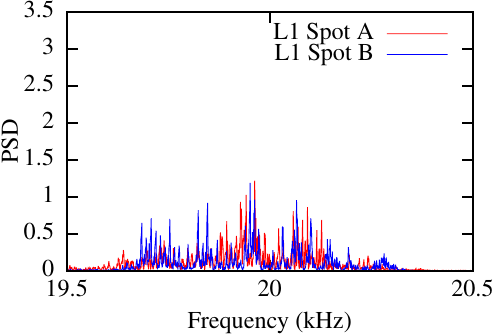}
    \label{fig:l1s6s1}
  }
  \subfigure[Different rooms]
  {
    \includegraphics{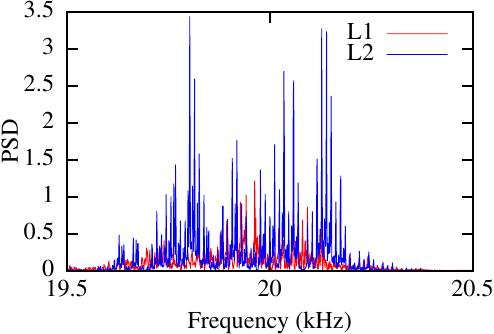}
    \label{fig:l1s6l2s1}
  }
  \caption{Frequency responses of rooms L1 and L2. (Please view the color version for better visibility.)}
  \label{fig:freq-response}
\end{figure}

The time-domain analysis shows the presence of indoor echos in response to single-tone chirps. We conjecture that different rooms have different frequency responses to the chirps. As the fast Fourier transform (FFT) needs $x$ seconds of data to generate a spectrum with a resolution of $1/x\,\text{Hz}$, the resolution of the spectrum based on the data in an echo data period of $97.5\,\text{ms}$ is $10.3\,\text{Hz}$ only. To improve the resolution, we concatenate the data in 40 echo data periods and then apply FFT to achieve a resolution of $0.26\,\text{Hz}$. Fig.~\ref{fig:l1s6} shows the power spectral densities (PSDs) in the frequency range of $[19.5, 20.5]\,\text{kHz}$ for the data collected at the same spot in room L1 at two different times, respectively. The PSDs remain stable over time. Fig.~\ref{fig:l1s6s1} shows the PSDs for two different spots in room L1. We can see that they are also similar. Fig.~\ref{fig:l1s6l2s1} shows the PSDs for the data collected from rooms L1 and L2, respectively. Although L1 and L2 have the same size (cf.~Fig.~\ref{fig:lab-floor-plan}), the materials in them may have different acoustic absorption rates. In Fig.~\ref{fig:l1s6l2s1}, L2 has stronger echos than L1. Moreover, the peak frequencies of the L2's responses are quite different from L1's. The results in Fig.~\ref{fig:freq-response} show that the rooms L1 and L2, though with the same size, exhibit different frequency responses. This is indicative of the differentiability of the rooms based on their frequency responses to single-tone inaudible chirps.

However, a total of four seconds will be needed to collect data for concatenating 40 echo data periods. This will incur privacy concerns and increase computation overhead since a total of 172,000 data points need to be processed. It is desirable to minimize the audio recording time to mitigate the potential privacy concerns and reduce computation overhead. In this paper, we inquire the possibility of using an audio record collected during a single echo data period of $97.5\,\text{ms}$ to recognize a room, since we believe that in general no meaningful private information can be extracted by an inspection on a $97.5\,\text{ms}$ audio record. A possible approach is to apply FFT on the 4,300 data points in the echo data period to generate a PSD and use the $[19.5, 20.5]\,\text{kHz}$ band to recognize a room. In Section~\ref{sec:approach}, this short-time PSD is employed as a possible format of the raw data input to the deep room recognizer.



\subsubsection{Time-Frequency Analysis}
\label{subsubsec:time-frequency}

\begin{figure}
  \centering
  \subfigure[Room L1, Spot A]
  {
    \includegraphics{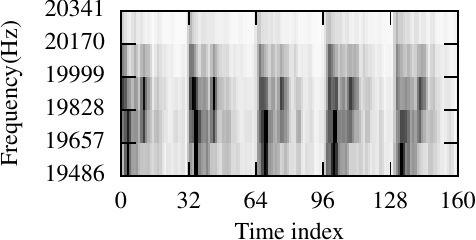}
    \label{fig:l1a}
  }
  \subfigure[Room L1, Spot B]
  {
    \includegraphics{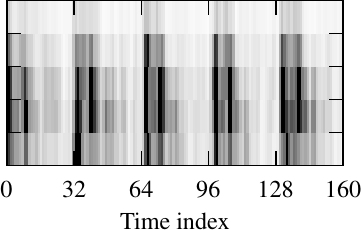}
    \label{fig:l1b}
  }
  \subfigure[Room L1, Spot C]
  {
    \includegraphics{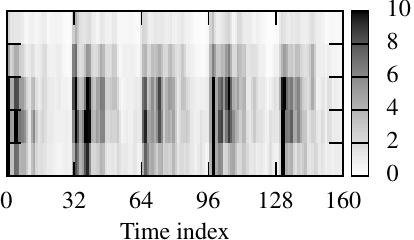}
    \label{fig:l1c}
  }
  \subfigure[Room L2, Spot A]
  {
    \includegraphics{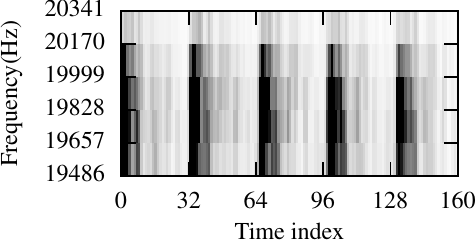}
    \label{fig:l2a}
  }
  \subfigure[Room L2, Spot B]
  {
    \includegraphics{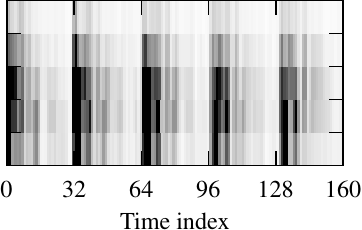}
    \label{fig:l2b}
  }
  \subfigure[Room L2, Spot C]
  {
    \includegraphics{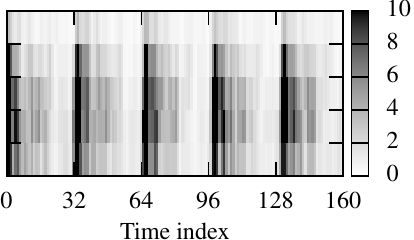}
    \label{fig:l2c}
  }
  \caption{Spectrograms at different spots in different rooms.}
  \label{fig:spectrograms}
\end{figure}

Our measurements in Section~\ref{subsubsec:time-domain} and \ref{subsubsec:frequency-domain} show that the bouncebacks of the echos form a process over time. Moreover, the tested rooms L1 and L2 exhibit distinct frequency responses. Thus, we investigate whether the spectrogram, a time-frequency representation of the raw data, can characterize a room effectively. Specifically, we apply 256-point Hann windows, with 128 points of overlap between two neighbor windows, to generate a total of 32 data blocks from the 4,300 data points in the echo data period. We note that the Hann windowing suppresses the side lobes of the PSD computed by the short-time FFT. The concatenation of all blocks' PSDs over time forms a spectrogram. As each PSD has five points only in the frequency range of interest, i.e., $[19.5,20.5]\,\text{kHz}$, the spectrogram that we use is a monochrome image with a dimension of 32 (time) $\times$ 5 (frequency). Fig.~\ref{fig:l1a} shows five concatenated spectrograms corresponding to five chirps when the phone is placed at spot A in room L1. We can see that the spectrograms exhibit similar patterns. Fig.~\ref{fig:l1b} and \ref{fig:l1c} show the results for two other spots, B and C, in room L1. Each spot has similar spectrograms. Moreover, we can observe some differences among the spectrograms at the three spots. Fig.~\ref{fig:l2a}, \ref{fig:l2b}, and \ref{fig:l2c} show the spectrograms at three spots in room L2. Although the rooms L1 and L2 have the same size and the same furniture (cf.~Fig.~\ref{fig:lab-floor-plan}), their spectrograms show perceptible differences. Specifically, each spectrogram in the room L1 consists of two or more disjunct segments in time, whereas each spectrogram in the room L2 is a more unified segment. This is because the two rooms' responses to the chirp have different time-domain characteristics.

From the results shown in Fig.~\ref{fig:spectrograms}, the tested rooms L1 and L2, though with the same size and furniture, show distinct echo spectrograms in response to single-tone chirps. This observation suggests that it is possible to recognize a room using a short audio record.
However, the spectrograms at different spots in the same room also exhibit some differences. Therefore, it is interesting to develop a classifier that can differentiate rooms while remaining insensitive to the small differences among different spots in the same room.



\section{Deep Room Recognition}
\label{sec:approach}

Based on the observations in Section~\ref{sec:measurement}, this section presents the design of the deep model for room recognition. Section~\ref{subsec:background} introduces the background  of deep learning and states the research problem. Section~\ref{subsec:deep-model-design} presents a set of preliminary trace-driven experiments to evaluate the performance of Deep Neural Network (DNN) and Convolutional Neural Network (CNN). The results show that CNN outperforms DNN. Section~\ref{subsec:hyper-settings} designs the hyperparameters of the CNN through a set of experiments.




\subsection{Background and Problem Statement}
\label{subsec:background}


The performance of the traditional classification algorithms such as Bayes classifiers and SVM highly depends on the effectiveness of the designed feature.
The feature design is often through a manual {\em feature engineering} process that exhaustively examines various popular dimension reduction techniques. For instance, in the audio processing landscape, handcrafted MFCC \cite{zheng2001comparison} and Perceptual Linear Predictive (PLP) coefficients \cite{hermansky1990perceptual} are widely used as the basis for audio feature design. 
The emerging deep learning methods \cite{lecun2015deep} replace the manual feature engineering process with automated feature learning. Thus, deep learning can substantially simplify the design of the pattern recognition system. More importantly, fed with sufficient training data, deep learning algorithms can effectively capture the intricate representations for feature detection or classification, thus yielding higher recognition accuracy over the traditional classification algorithms. This has been evidenced by the recent successful applications of deep learning \cite{lecun2015deep}. 

DNN and CNN are two types of deep models that are widely used for audio sensing (e.g., speech recognition) and image classification, respectively. A DNN consists of a series of fully connected layers with each layer comprised by a collection of neurons (or units). The data to be classified initialize the values of the input layer neurons. Multiple hidden layers follow the input layer. The yield of a hidden layer neuron is the output of an activation function that takes the weighted sum of the outputs of all the previous layer's neurons as input. Thus, a hidden layer neuron is not connected with any other neuron in the same layer, but is fully connected to all neurons in the previous layer. In the last layer, i.e., the output layer, the neuron giving the largest value indicates the class of the input data. The training algorithm determines the weights and biases of the neurons to best fit the DNN to the labeled training data. Different from DNN that is often used to classify 1-dimensional data, CNN is good at capturing local patterns in data with higher dimensions that largely determine the class of the data. CNN consists of one or more convolutional, pooling, and fully connected (or dense) layers that respectively searches for the local patterns (i.e., feature extraction), increases the extracted feature's robustness to data translations (e.g., rotation), and votes the classification result. The parameters of the neurons in the convolutional and dense layers are determined in the training process, whereas the pooling layers have no parameters to be trained.

The key question we ask in this paper is whether we can recognize a room based on its acoustic response to a $2\,\text{ms}$ $20\,\text{kHz}$ single-tone chirp. Due to the limited time and frequency spans of the chirp, the response may carry limited information about the room. To address this challenge, we apply deep learning that can capture the differences deeply embedded in the raw data of different classes. To this end, we need to design the appropriate format of the raw data, the deep model, and the model's hyperparameters. These issues will be addressed in Sections~\ref{subsec:deep-model-design} and \ref{subsec:hyper-settings}.



\subsection{Design of Raw Data Format and Deep Model}
\label{subsec:deep-model-design}

\subsubsection{Candidate Raw Data Formats}

As shown in Sections~\ref{subsubsec:frequency-domain} and \ref{subsubsec:time-frequency}, both the frequency-domain and time-frequency representations of the echo data can be indicative of the rooms' differences. Thus, the PSD and spectrogram are two possible raw data formats for deep learning.
To avoid privacy concern, we apply FFT on the 4,300 data points in the echo data period to generate a short-time PSD, rather than concatenating 40 echo data periods as in Section~\ref{subsubsec:frequency-domain}. Then, we only use the 147 points in the $[19.5, 20.5]\,\text{kHz}$ band of the short-time PSD as the input data to a deep model. Following the approach in Section~\ref{subsubsec:time-frequency}, the spectrogram for the 4,300 data points has a dimension of 32 (time) $\times$ 5 (frequency). Thus, the data volumes of the short-time PSD segment and the spectrogram are similar (i.e., 147 and 160).


\subsubsection{Candidate Deep Models}
\label{subsubsec:cnn-design}


We implement DNN and CNN using Python based on Google TensorFlow \cite{abadi2016tensorflow}. The structures and hyperparameters of the deep models are designed as follows. In Section~\ref{subsec:hyper-settings}, we will conduct extensive experiments to optimize the hyperparameters.

DNN admits one-dimensional inputs only. Thus, the PSD segment can be used directly with the DNN. For the spectrogram, we flatten it as a vector with a length of 160 and then use the vector as the input to the DNN. The DNN has two hidden layers with each layer comprised by 256 rectified linear units (ReLUs). Suppose there are $K$ rooms in the training dataset. The output layer consists of $K$ ReLUs that correspond to the $K$ target classes. Note that the training of ReLU-based neural networks is often several times faster than that of the traditional tanh-based and sigmoid-based networks \cite{krizhevsky2012imagenet,lane2015deepear}.

\begin{figure}
  \includegraphics{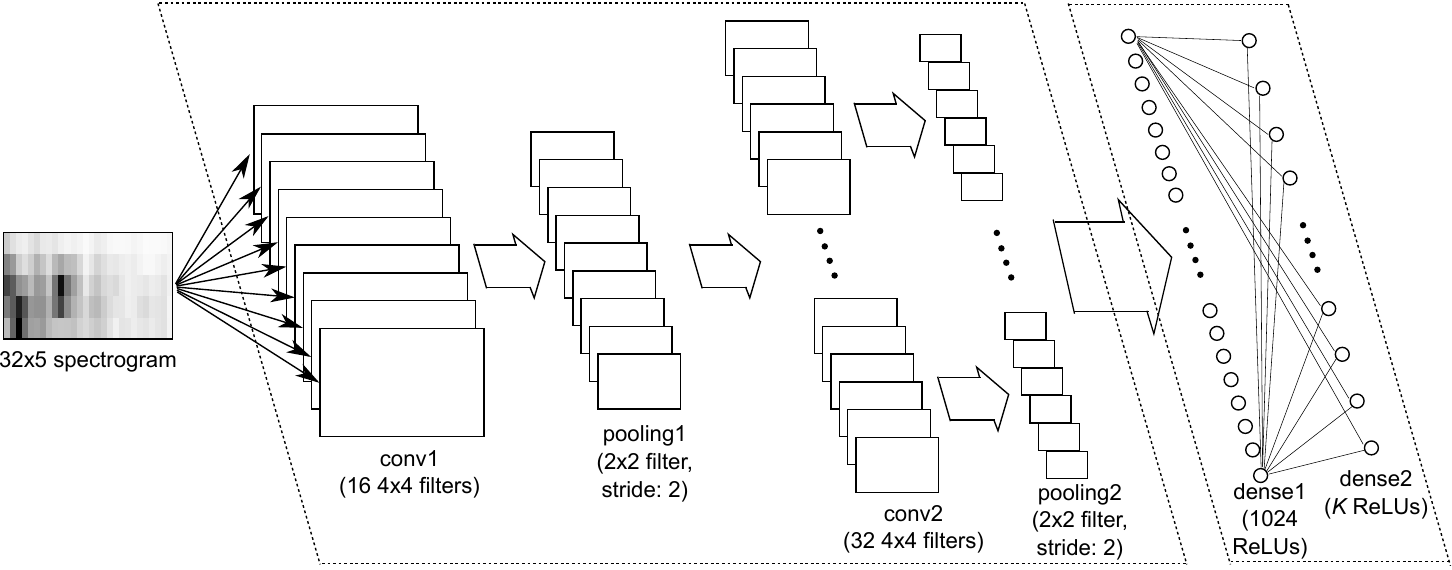}
  \caption{CNN for room recognition.}
  \label{fig:CNN}
\end{figure}

CNN can admit high-dimensional inputs. In what follows, we describe the design of the CNN that takes the two-dimensional spectrogram as the input. As illustrated in Fig.~\ref{fig:CNN}, the CNN consists of the following layers: {\texttt{conv1}, \texttt{pooling1}, \texttt{conv2}, \texttt{pooling2}, \texttt{dense1}, and \texttt{dense2}. The first four layers extract features from the input data by applying sets of filters that preserve the spatial structure of the input data. The dense layers are equivalent to a DNN that performs classification. These layers are briefly explained as follows.

\begin{itemize}
\item The \texttt{conv1} layer applies 16 $4 \times 4$ convolution filters to the $32 \times 5$ spectrogram. We add zero padding to the edges of the input image such that the filtered image has the same dimension as the input image.
A filter is slid over the input image by one pixel each time, yielding a single value in the output image that is computed by an element-wise arithmetic operator. Thus, the \texttt{conv1} layer generates 16 output images.
It further applies the ReLU to rectify the negative pixel values in the 16 output images to zero. This introduces non-linearity that is generally needed by neural networks.
\item The \texttt{pooling1} layer performs the max pooling with a $2\times 2$ filter and a stride of two to each output image of the \texttt{conv1} layer. Specifically, a $2 \times 2$ window is slid over the image by two pixels each time, yielding the maximum pixel value in the covered $2 \times 2$ subregion as a pixel of the output image. Thus, the output image has a dimension of $16 \times 2$. As pooling downsizes the feature image, it can control overfitting effectively. Moreover, as it generates summary of each subregion, it increases the CNN's robustness to small distortions in the input image.
\item The \texttt{conv2} and \texttt{pooling2} layers perform similar operations as the \texttt{conv1} and \texttt{pooling1} layers, respectively. The \texttt{conv2} layer applies 32 $4 \times 4$ filters with ReLU rectification to generate 32 $14 \times 2$ images. Then, the \texttt{pooling2} applies max pooling and generates 32 $8 \times 1$ images. These images are flattened and concatenated to form a feature vector with a length of 256 ($8 \times 1 \times 32 $).
\item The \texttt{dense1} layer consists of 1,024 ReLUs. The feature vector from the \texttt{pooling2} layer is fully connected to all these ReLUs.
We apply dropout regulation to avoid overfitting and improve the CNN's performance. Specifically, we apply a dropout rate of 0.4, i.e., 40\% of the input features will be randomly abandoned during the training process.
\item The \texttt{dense2} layer consists of $K$ ReLUs that corresponds to the $K$ target rooms. For an input spectrogram, the classification result corresponds to the maximum value among the $K$ ReLUs.
\end{itemize}


The above CNN design is for two-dimensional spectrogram. To use the one-dimensional PSD segment with the CNN, we make the following minor changes to the above design:
\begin{itemize}
  \item The size of the convolution filters in the \texttt{conv1} and \texttt{conv2} layers is changed to $1 \times 4$.
  \item The size of the filters in the \texttt{pooling1} and \texttt{pooling2} layers is changed to $1 \times 2$.
\end{itemize}


\subsubsection{Model Training}

The training process of the DNN and CNN is as follows. We initialize the neural network's parameters randomly. For each step of the training, a mini-batch of randomly selected 100 training samples is fed to the neural network. The network performs the forward propagation and computes the cross entropy between the output of the \texttt{dense2} layer and the one-hot vector formed by the labels of the 100 training samples. The cross entropy is often used as the loss metric to assess the quality of multiclass classification. Based on the cross entropy, stochastic gradient descent is employed to optimize the neural network's parameters over many training steps. We set the learning rate to be 0.001. The training can be stopped when the number of the training steps reaches a predefined value or the loss metric does not reduce anymore.

\subsubsection{Preliminary Results.}
\label{subsubsec:deep-model-prel}

To design the deep room recognition algorithm, we collect 22,000 samples from 22 rooms in three homes, an office, and the lab shown in Fig.~\ref{fig:lab-floor-plan}. Each sample is a $100\,\text{ms}$ audio record. We split the data set into three parts: training set, validation set, and testing set. Among 1,000 samples collected from each room, 500, 250, and 250 samples are used as training, validation, and testing data, respectively.
  In each training-validation epoch, the deep model parameters are tuned by the stochastic gradient descent based on the training data and the average classification accuracy is computed using the validation data. The epoch is repeated until the average classification error no longer decreases substantially. The testing data set is used to measure the average classification accuracy of the trained deep model. The testing data is previously unseen by the training-validation phase. In the rest of this paper, all accuracy results are the average classification accuracy measured using the testing data.
On a workstation computer with an Intel Xeon E5-1650 processor, 16GB main memory, and a GeForce GTX 1080 Ti graphics processing unit (GPU), the training of the spectrogram-based DNN and CNN achieves the peak validation accuracy after about two minutes.





\begin{table}
  \caption{Average accuracy of four possible designs in classifying 22 rooms.}
  \label{table:design}
  \begin{tabular}{c|cc}
    \Xhline{1pt}
    & PSD segment & Spectrogram \\
    \hline
    DNN & 19\% & 80\% \\
    CNN & 33\% & 99\% \\
    \Xhline{1pt}
  \end{tabular}
\end{table}

Based on the testing data, the average accuracy of the four possible designs in classifying the 22 rooms is shown in Table~\ref{table:design}. From the results, we can see that, although both the PSD segment and the spectrogram represent the same raw data, the deep models fed with the spectrogram give much higher classification accuracy. This is because the distinction in the time dimension among the rooms' responses to the chirps is more salient than the distinction in the frequency dimension. Thus, although the spectrogram has a much lower frequency resolution than the PSD segment (i.e., 5 points vs. 147 points), the spectrogram is more effective in expressing the response of a room. Based on the spectrogram, DNN and CNN achieve 80\% and 99\% classification accuracy, respectively.
Although in this preliminary test we do not extensively optimize the hyperparameters of the two deep models, the test result is consistent with the common understanding that CNN is better in classifying images.
Thus, in the rest of this paper, we choose the combination of spectrogram and CNN.


\subsection{Hyperparameter Settings}
\label{subsec:hyper-settings}

The results in Section~\ref{subsec:deep-model-design} show that CNN is an appropriate deep model for room recognition.
This section presents our experiments to decide the settings of the following hyperparameters: the number of convolutional layers, the presence of pooling layers, the number of filters in the convolutional layers, and the sizes of the filters. In each experiment, we vary a single hyperparameter and keep others unchanged.
For each setting, we train and test the CNN using the training-validation and testing samples collected from the 22 rooms as described in Section~\ref{subsubsec:deep-model-prel}.


\subsubsection{The Number of Convolutional Layers}

\begin{table}[]
	\centering
	\caption{The configurations of the tested CNNs with one to five convolutional layers.}
	\label{table:DRR-configuration}
	\begin{tabular}{|lllllll|}
		\hline
		\multicolumn{1}{|l|}{CNN-A} & \multicolumn{1}{l|}{CNN-B} & \multicolumn{1}{l|}{CNN-C} & \multicolumn{1}{l|}{CNN-D} & \multicolumn{1}{l|}{CNN-E} & \multicolumn{1}{l|}{CNN-F} & CNN-G \\ 
		\multicolumn{1}{|l|}{1 conv layer} & \multicolumn{1}{l|}{2 conv layers} & \multicolumn{1}{l|}{2 conv layers} & \multicolumn{1}{l|}{2 conv layers} & \multicolumn{1}{l|}{3 conv layers} & \multicolumn{1}{l|}{4 conv layers} & 5 conv layers \\ \hline
		&                       &                       &            $5 \times 32$ images           &                       &                       &  \\
		\hline
		\multicolumn{1}{|l|}{conv4-16} & \multicolumn{1}{l|}{conv4-16} & \multicolumn{1}{l|}{conv4-16} & \multicolumn{1}{l|}{conv4-32} & \multicolumn{1}{l|}{conv4-16} & \multicolumn{1}{l|}{conv4-16} & conv4-16 \\ \hline
		&                       &                       &          max pooling        &                       &                       &  \\ \hline
		\multicolumn{1}{|l|}{n.a.} & \multicolumn{1}{l|}{conv4-16} & \multicolumn{1}{l|}{conv4-32} & \multicolumn{1}{l|}{conv4-32} & \multicolumn{1}{l|}{conv4-32} & \multicolumn{1}{l|}{conv4-32} & conv4-32 \\ \hline
		\multicolumn{1}{|l|}{n.a.} &                       &                       &          max pooling           &                       &                       &  \\ \hline
		\multicolumn{1}{|l|}{n.a.} & \multicolumn{1}{l|}{n.a.} & \multicolumn{1}{l|}{n.a.} & \multicolumn{1}{l|}{n.a.} & \multicolumn{1}{l|}{conv4-64} & \multicolumn{1}{l|}{conv4-64} & conv4-64 \\
		\multicolumn{1}{|l|}{} & \multicolumn{1}{l|}{} & \multicolumn{1}{l|}{} & \multicolumn{1}{l|}{} & \multicolumn{1}{l|}{} & \multicolumn{1}{l|}{conv4-128} & conv4-128 \\
		\multicolumn{1}{|l|}{} & \multicolumn{1}{l|}{} & \multicolumn{1}{l|}{} & \multicolumn{1}{l|}{} & \multicolumn{1}{l|}{} & \multicolumn{1}{l|}{} & conv4-256 \\ \hline
		&                       &                       &            dense-1024          &                       &                       &  \\ \hline
		&                       &                       &            dense-$K$           &                       &                       &  \\ \hline
		&                       &                       &             softmax          &                       &                       &  \\ \hline
	\end{tabular}
	\footnotesize conv$x$-$n$ represents a total of $n$ $x \times x$ convolution filters; dense-$n$ represents a dense layer with $n$ ReLUs.
      \end{table}
      
First, we study the impact of the number of convolutional layers on the classification accuracy. We follow the test methodology used in the design of the VGG net \cite{2014arXiv1409.1556S}. We test a total of seven CNNs, named from CNN-A to CNN-G, with one to five convolutional layers. The configurations of these CNNs are illustrated in Table~\ref{table:DRR-configuration}, in which ``conv$x$-$y$'' means a total of $y$ convolutional filters with size of $x \times x$. All convolutional filters use a stride of one pixel and zero padding.
Max pooling with window size of $2 \times 2$ and stride of two is applied after some of the convolutional layers.
Note that we can apply at most two max pooling layers, since after that the image size reduces to $1 \times 8$.
All tested CNNs have two dense layers. The first has 1024 ReLUs, whereas the second consists $K$-way classification channels for the last soft-max layer that gives the final classification result.

Table~\ref{table:design-nlayer} shows the total number of the neurons' parameters, training time on the aforementioned workstation computer, and classification accuracy of different CNNs illustrated in Table~\ref{table:DRR-configuration}.
We note that, as CNN-A has one max pooling layer only and pooling can reduce the size of the images going through the network, CNN-A contains more parameters than CNN-B, C, and D that have two convolutional and pooling layers.
We can see that the training time increases with the number of parameters, but the accuracy does not. The CNN-C with two convolutional layers achieves the highest accuracy among the tested CNNs. Moreover, the accuracy of the CNNs with two convolutional layers (i.e., CNN-B, C, and D) is generally higher than other tested CNNs.
We note that more layers or more parameters unnecessarily lead to better accuracy due to potential overfitting. From the results in Table~\ref{table:design-nlayer}, we adopt two convolutional layers in the rest of this paper.



\begin{table}
	\begin{minipage}{\columnwidth}
		\begin{center}
			\caption{Training time and accuracy of various CNNs.}
			\label{table:design-nlayer}
			\begin{tabular}{lccccccc}
				\toprule
				CNN- & A & B & C & D & E & F & G\\
				\midrule
				The number of parameters (million) & 0.48 & 0.14 & 0.26 & 0.26 & 0.52 & 1.11 & 2.55\\
				Training time (second) & 115 & 112 & 119 & 124 & 134 & 158 & 203 \\
				Accuracy (\%) & 93.6 & 95.7 & 99.7 & 95.1 & 90.2 & 74.4 & 58.4\\
				\bottomrule
			\end{tabular}
		\end{center}
	\end{minipage}
\end{table}

\subsubsection{Presence of Pooling Layers}

As discussed in Section~\ref{subsec:background}, the pooling layers have no parameters to be trained. But their presence can be decided. As the main function of pooling is to reduce the amount of parameters and computation time, as well as increase the CNN's robustness to data translations, the pooling layers can be omitted for input images with relatively small dimensions \cite{Li:2016:DLR:2994551.2994569}.
Our tests show that, by omitting the pooling lays, the accuracy of the CNN increases from 99.7\% to 99.9\%, probably due to the small dimensions of the spectrogram. However, the omission results in tripled training time. As long training times are undesirable when our room recognition system runs in a participatory learning mode (cf.~Section~\ref{sec:rr}), the 0.2\% accuracy gain is not worth. Thus, we retain the pooling layers.


\subsubsection{The Number and Size of Filters}

\begin{table}
  \begin{center}
    \caption{Accuracy under various numbers of filters in the two convolutional layers.}
    \label{table:design-nfilter}
    \begin{tabular}{lcccc}
      \toprule
      The number of filters & (8, 16) & (16, 32) & (32, 64) & (64, 128) \\
      \midrule
      Accuracy (\%) & 92.1 & 99.7 & 96.4 & 94.8\\
      \bottomrule
    \end{tabular}
  \end{center}
  \centering
  \footnotesize ($x$, $y$): $x$ filters in \texttt{conv1}, $y$ filters in \texttt{conv2}. The size of the filters is $4 \times 4$.
\end{table}

We vary the numbers of the filters in the two convolutional layers by looping through the powers of 2 from 16 to 256. Table~\ref{table:design-nfilter} shows the resulting accuracy. Note that \texttt{conv2} has doubled filters compared with \texttt{conv1}. This is a typical setting adopted in many CNNs (e.g., VGG net \cite{2014arXiv1409.1556S} and DenseNet \cite{huang2016densely}).
When \texttt{conv1} and \texttt{conv2} have 16 and 32 filters, the CNN gives the highest accuracy. This is because more filters lead to more parameters, but unnecessarily better accuracy due to potential overfitting. Thus, we adopt 16 and 32 filters for the two layers in our approach.

We also test the impact of the filter size on the CNN's accuracy by varying the size from $2 \times 2$ to $5 \times 5$. Table~\ref{table:design-filtersize} shows the resulting accuracy. Similarly, the accuracy is concave to the filter size due to potential overfitting. In particular, the filter size $4 \times 4$ gives the highest accuracy.


\begin{table}
  \centering
  \begin{minipage}{.45\columnwidth}
    \begin{center}
      \caption{Accuracy under various filter sizes.}
      \label{table:design-filtersize}
      \begin{tabular}{lcccc}
        \toprule
        Filter size (pixel) & $ 2 \times 2 $ & $ 3 \times 3 $ & $ 4 \times 4 $& $ 5 \times 5 $\\
        \midrule
        Accuracy (\%) & 96.9 & 99.2 & 99.7 & 96.8\\
        \bottomrule
      \end{tabular}
    \end{center}
    \centering
    \footnotesize 16 and 32 filters in \texttt{conv1} and \texttt{conv2}, respectively.
  \end{minipage}
  \begin{minipage}{.54\columnwidth}
    \begin{center}
      \caption{Accuracy under various depth of dense layers.}
      \label{table:design-ndenselayer}
      \begin{tabular}{lccc}
        \toprule
        Number of dense layers& 2 & 3 & 4 \\
        \midrule
        Number of parameters (million)& 0.26 & 1.3 & 2.3 \\
        Training time (minute)& 2 & 3 & 4 \\
        Accuracy (\%)& 99.7 & 98.9 & 99.1 \\
        \bottomrule
      \end{tabular}
    \end{center}
  \end{minipage}
\end{table}

\subsubsection{The Number of Dense Layers}

We vary the number of dense layers from one to three. The last dense layer has $K$ ReLUs, whereas each of prior dense layers has 1,024 ReLUs. Each dense layer adopts dropout. Table~\ref{table:design-ndenselayer} shows the resulting number of neurons' parameters, training time, and accuracy. We can see that the configuration of two dense layers as illustrated in Table~\ref{table:DRR-configuration} needs the least training time and gives the highest accuracy. Thus, we adopt two dense layers.


\subsubsection{Summary and Discussions}
From the above results, the hyperparameter settings for CNN adopted in Section~\ref{subsubsec:cnn-design}, i.e., CNN-C shown in Table~\ref{table:design-nlayer}, are preferable. Thus, we design our room recognition approach based on CNN-C. We now discuss two issues relevant to hyperparameter design. First, although the hyperparameter settings are designed based on the data collected from 22 rooms, in Section~\ref{sec:eval}, we will evaluate the performance of CNN-C in classifying more rooms and other types of spaces such as location spots in museums. The results show that CNN-C yields excellent/good classification accuracy in all evaluation cases. Second, more systematic techniques such as grid search and AutoML \cite{automl} can be employed to further optimize the hyperparameter settings. However, since our design experiments have achieved an accuracy of 99.7\%, the accuracy improvement by these hyperparameter optimization techniques will not be substantial. We leave the integration of these techniques when massive training data is available to our future work.




\section{Deep Room Recognition Cloud Service}
\label{sec:rr}

\subsection{System Overview}
\label{subsec:sys-overview}

Based on the results in Section~\ref{sec:approach}, we design {\em RoomRecognize}, a cloud service for room recognition, and its mobile client library. RoomRecognize, running on a cloud server with GPU support, classifies the echo data sent from a mobile client. With the mobile client library, the application developer can readily integrate the cloud service into mobile applications that need room recognition services. Fig.~\ref{fig:RoomRecognize_Arch} shows the architectures of RoomRecognize and its client library. In particular, we design RoomRecognize to support a {\em participatory learning} mode, in which the CNN is retrained when a mobile client uploads labeled training samples. Fig.~\ref{fig:Participatory_Workflow} shows the workflow of the participatory learning mode. First, the client collects training samples in a room and uploads to the server. Then, the server will run the current CNN using the training samples and return a list of the most probable room labels to the client. The user of the client can check the list. If the current room is not in the list, the user can create a new label and trigger the server to retrain the CNN; otherwise, the uploaded training samples should be labeled using an existing room label before being used for future CNN retraining. This design helps prevent multiple different labels defined by the users for the training data collected from the same room.


We note that existing studies \cite{lane2015deepear,yao2017compressing} have shown that smartphones and even lower-end Internet of Things platforms can run deep models. In RoomRecognize, as the transmission of an audio record of 0.1 seconds causes little overhead to today's communication networks, we choose to run the trained CNN at the cloud server. This design choice also avoids the complications in synchronizing the latest deep model to each client in the participatory learning mode.

\begin{figure}
  \begin{minipage}[t]{.55\textwidth}
    \includegraphics[width=\textwidth]{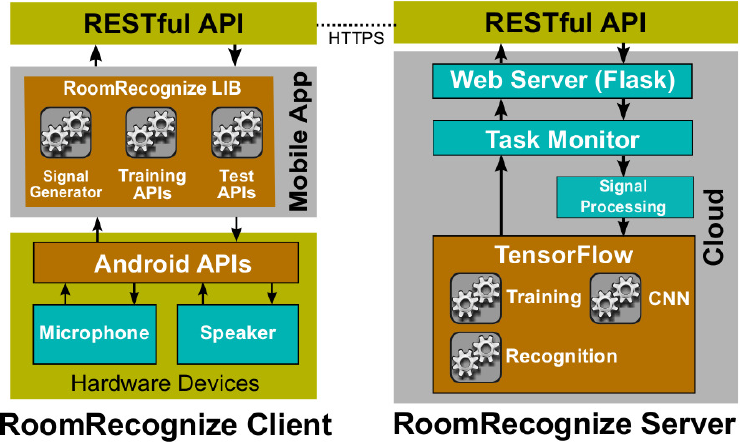}
    \caption{Architecture of RoomRecognize.}
    \label{fig:RoomRecognize_Arch}    
  \end{minipage}
  \hspace{.02\textwidth}
  \begin{minipage}[t]{.42\textwidth}
    \includegraphics[width=\textwidth]{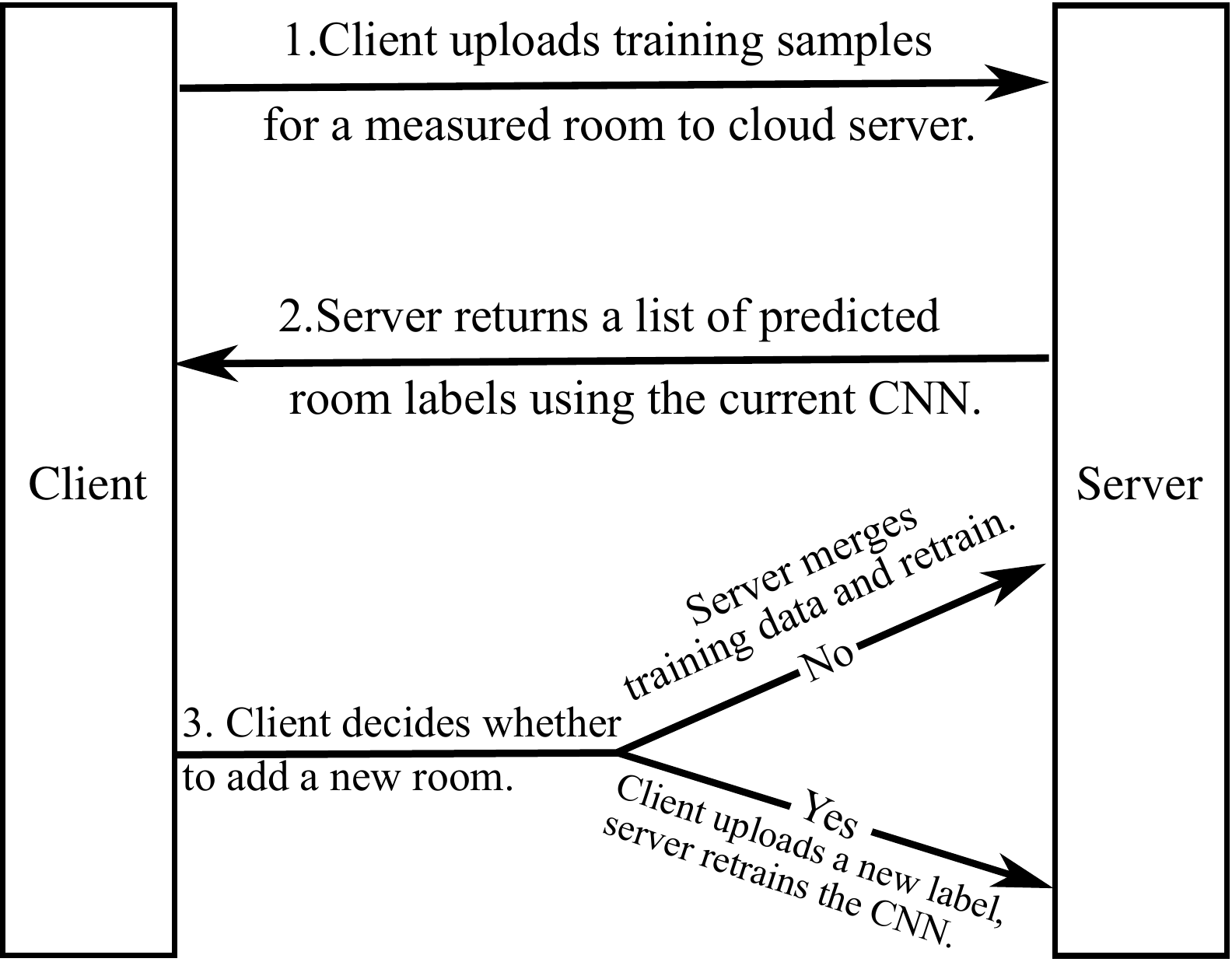}
    \caption{Workflow of participatory learning.}
    \label{fig:Participatory_Workflow}
  \end{minipage}
\end{figure}


\subsection{RoomRecognize Service}

We build RoomRecognize based on Flask \cite{flask}, a Python-based micro web framework. Thus, we can easily integrate the Python-based TensorFlow into Flask. We use the Flask-RESTful extension to develop a set of representational state transfer (RESTful) APIs over HTTPS. Note that RESTful APIs can largely simplify the interoperations between the cloud service and the non-browser-based client programs. Through these APIs, the client can upload an audio record of 0.1 seconds and obtain room recognition result, or training samples of multiple labeled audio records. The functionality of these APIs will be further explained in Section~\ref{subsec:client-lib}.
The {\em Signal Processing} module shown in Fig.~\ref{fig:RoomRecognize_Arch} extracts the spectrograms from the received audio records. The {\em Task Monitor} manages the training of the CNN with newly received training data in the participatory learning mode. As the CNN training is performed by a separate process, it will not block the room recognition service. Once the training completes, the deep model is updated. All modules of RoomRecognize are implemented using Python. In this work, we deployed the RoomRecognize service to a server equipped with an Intel Core i7-6850K CPU, 64GB main memory, a Quadro P5000 and two GeForce GTX 1080 Ti GPUs.


\subsection{RoomRecognize Client Library}
\label{subsec:client-lib}

We design an Android client library in Java to wrap the RESTful APIs provided by the RoomRecognize cloud service. A similar library can be designed for iOS. The library provides the following methods:



\begin{itemize}
  \item \texttt{EmitRecord(mode)}: This method uses the phone's loudspeaker to transmit single-tone chirps and microphone to capture the echos. The \texttt{mode} can be either {\em recognition} or {\em training}. In the recognition mode, the phone emits a chirp only and records audio for 0.1 seconds. In the training mode, the phone repeats the above emit-record process for 500 times over 50 seconds. This is because that our evaluation in Section~\ref{sec:eval} shows that 500 training samples are sufficient to characterize a room. This \texttt{EmitRecord} method returns the recorded data to the user program.
  \item \texttt{UploadData(mode)}: In the recognition mode, this method uploads the audio record to the cloud service and obtains the recognition result of a single room label. In the training mode, it uploads the training samples and receives a list of at most five predicted rooms computed by the cloud service that have the highest neuron values in the CNN's \texttt{dense2} layer. In other words, these five predicted rooms are those in the server's training database that best match the currently measured room.
  \item \texttt{UploadLabel(label)}: This method should be used in the training mode only. The client can use this method to notify the server the label of the currently measured room. The label can be one within the list returned by the \texttt{UploadData()} method, or a new label. For the former case, the server will merge the training samples contributed by this client using the \texttt{UploadData()} method with the existing training samples from the same room; for the latter case, the server will create a new class and increase the $K$ value by one. Thereafter, the server's Task Monitor will trigger a retraining.
\end{itemize}

Note that, in our design, we separate the processes of uploading training data and label. This allows the application developer to easily deal with a situation where the end user contributes training data collected from a room that has been covered by the current training database. Specifically, the client program prompts the end user to check whether the current room is within the list returned by the \texttt{UploadData} method and then uploads the label based on the user's choice.

\section{Performance Evaluation}
\label{sec:eval}




Using the client library described in Section~\ref{sec:rr}, we have implemented an Android App to use the RoomRecognize service. In this section, we conduct a set of experiments using the Android App to evaluate the performance of RoomRecognize.


\subsection{Evaluation Methodology}
\label{subsec:eval-method}

Section~\ref{subsec:batphone-test} has shown the low performance of passive acoustic sensing, and its susceptibility to interfering sounds. In this section, we only compare RoomRecognize with other room recognition approaches that are based on active acoustic sensing. As discussed in Section~\ref{sec:related}, active acoustic sensing has been applied to recognize semantic locations \cite{kunze2007symbolic,fan2014public,tachikawa2016predicting} and remember predefined positions with centimeter resolution \cite{tung2015echotag}. However, these studies do not address room recognition. Thus, they are not comparable with RoomRecognize.

In our evaluation, we compare RoomRecognize with RoomSense \cite{rossi2013roomsense}. To the best of our knowledge, RoomSense is the only active acoustic sensing system for room recognition. RoomSense employs the maximum length sequence (MLS) measurement technique to generate chirps in a wide spectrum including the audible range, and then classifies a room using SVM based on MFCC features. We implement RoomSense by following the descriptions in \cite{rossi2013roomsense}. Specifically, we use Cool Edit Pro 2.1 to generate the MLS signal. We follow the measurement approach described in Section~\ref{subsubsec:measurement-setup}, except that the phone replays the MLS signal, to collect the rooms' responses. With the Python libraries \texttt{python\_speech\_features} \cite{python-speech} and \texttt{scikit-learn} \cite{scikit-learn}, we implement RoomSense's MFCC feature extraction and SVM-based classification. We also implement two variants of RoomSense. These two RoomSense variants emit the single-tone chirps described in Section~\ref{subsubsec:measurement-setup} instead of MLS signals, and then still apply the MFCC extraction and SVM classification pipeline to classify rooms. The first RoomSense variant, named {\em single-tone broadband-MFCC RoomSense}, sets the lowest and highest band edges for the MFCC extraction to be $0\,\text{kHz}$ and $22.05\,\text{kHz}$, same as the original setting of RoomSense. The second RoomSense variant, named {\em single-tone narrowband-MFCC RoomSense}, sets the two band edges to be $19.5\,\text{kHz}$ and $20.5\,\text{kHz}$. Thus, the single-tone narrowband-MFCC RoomSense and RoomRecognize extract acoustic features from the same frequency band. The comparisons among the RoomRecognize, the original RoomSense, and the two variants of RoomSense will provide insights into understanding the major factors contributing to RoomRecognize's high classification accuracy.

\begin{table}
  \centering
  \caption{ Descriptions of residential, office, teaching, and museum rooms involved in our evaluation.}
  \label{tab:description-rooms}
  \begin{tabular}{ccccccc}
    \toprule
    Room type & Number  & Size & \# of spots  & Wall & Floor & Ambient \\
    \multicolumn{2}{c}{} of rooms & ($\text{m}^2$) & per room & material & material & environment \\
    \midrule
    Bedroom & 9 & 10-20 & 2-3 & concrete & laminated/ceramic & generally quiet\\
    Living room & 2 & 25-30 & 3 & concrete & marble & slightly noisy \\
    Bathroom & 3 & 15-20 & 2-3 & ceramic & ceramic & quiet \\
    Kitchen & 2 & 10-15 & 2 & ceramic & ceramic & generally quiet \\
    Faculty office & 1 & 15 & 2 & concrete & ceramic & quiet \\
    Visitor office (L1) & 1 & 10 & 2 & concrete & ceramic & slightly noisy \\
    Visitor office (L2) & 1 & 10 & 2 & concrete  & ceramic & quiet \\
    Meeting room (L3) & 1 & 7 & 2 & concrete & ceramic & quiet \\
    Meeting room (L4) & 1 & 30 & 3 & concrete & ceramic & quiet \\
    Lab open area & 1 & 150 & 3 & concrete & ceramic & slightly noisy \\
    Teaching room & 10 & 40 & 1 & concrete & laminated & quiet \\
    Museum-A hall areas & 19 & 15-150 & 2-3 & concrete & ceramic & slightly noisy \\
    Museum-B hall areas & 15 & 30-100 & 2-3 & concrete & ceramic & noisy, crowded \\
    \bottomrule
  \end{tabular}
\end{table}

\begin{figure}
  \subfigure[Bedroom]
  {    
    \includegraphics[width=0.18\textwidth]{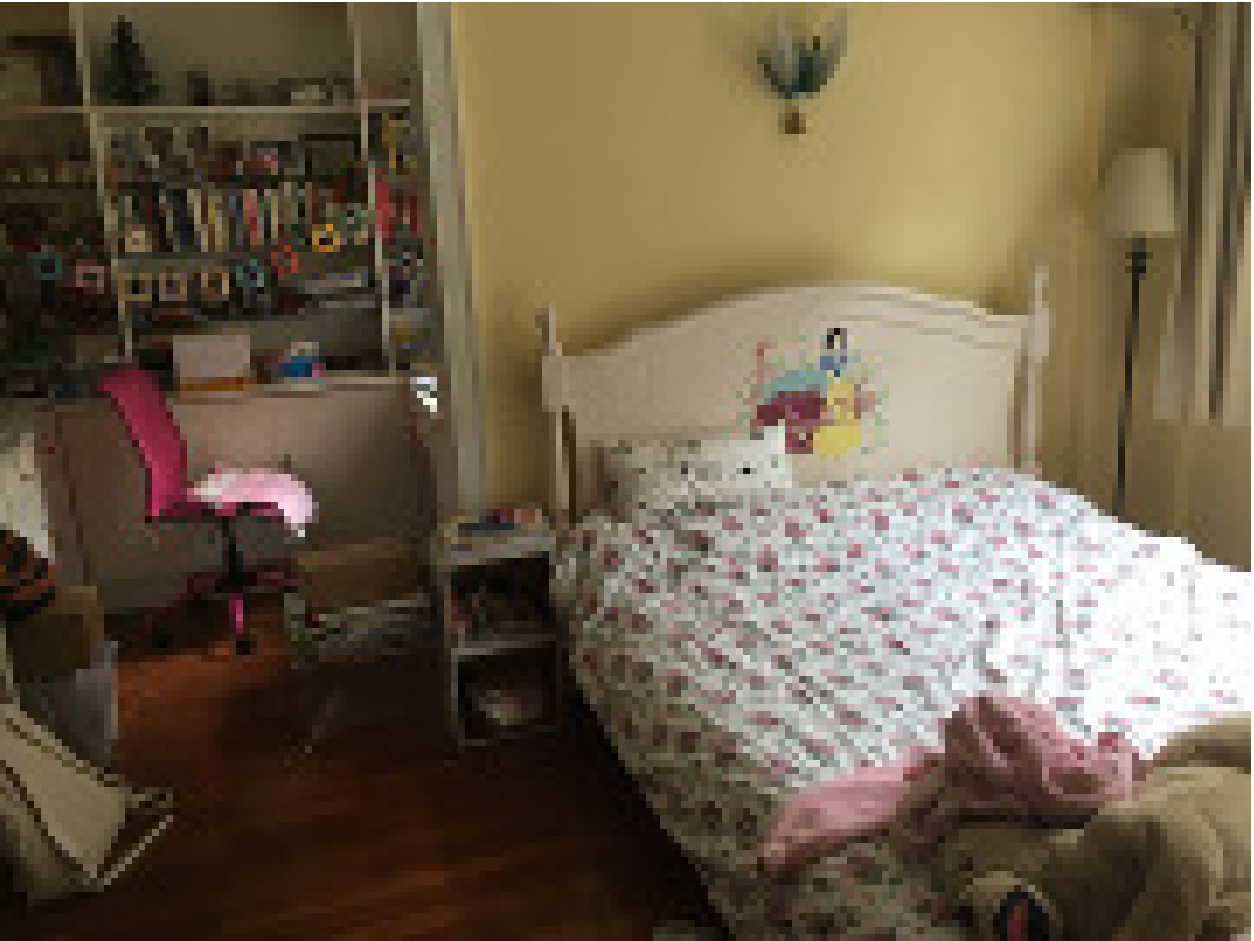}
  }
  \subfigure[Museum hall]
  {    
    \includegraphics[width=0.18\textwidth]{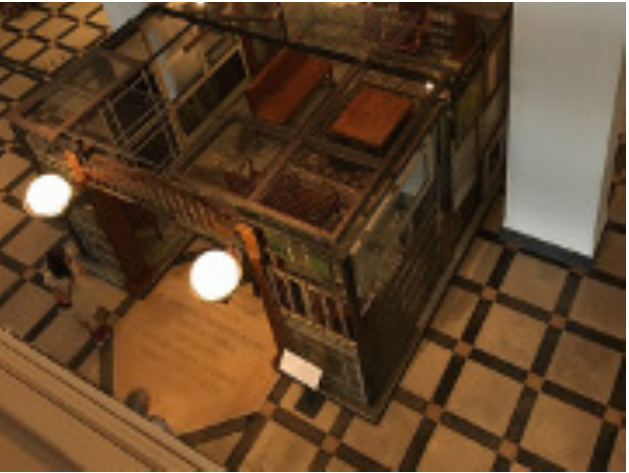}
  }
  \subfigure[Visitor office L1]
  {    
    \includegraphics[width=0.18\textwidth]{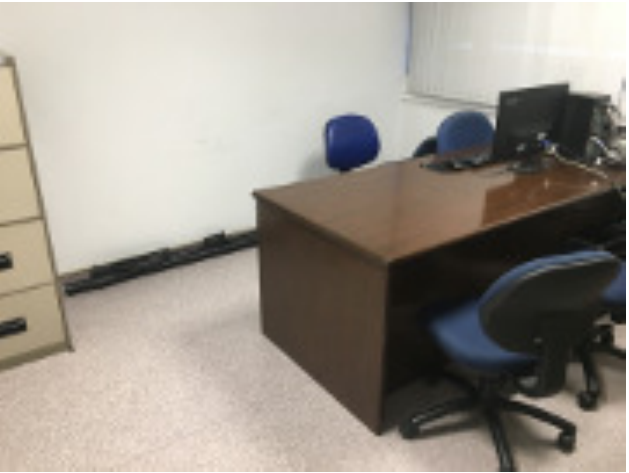}
  }
  \subfigure[Lab open area]
  {    
    \includegraphics[width=0.18\textwidth]{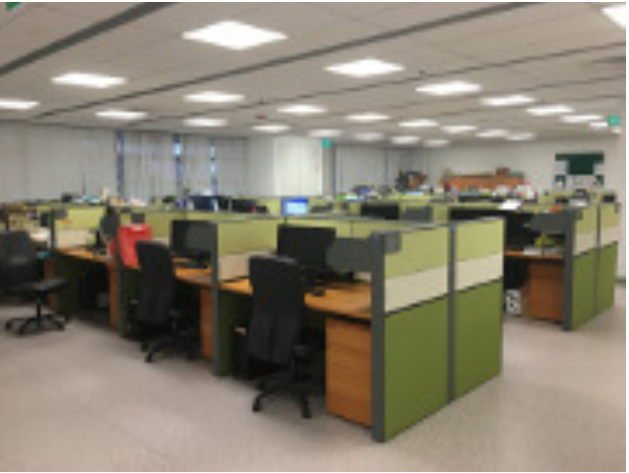}
  }
  \subfigure[Meeting room L4]
  {    
    \includegraphics[width=0.18\textwidth]{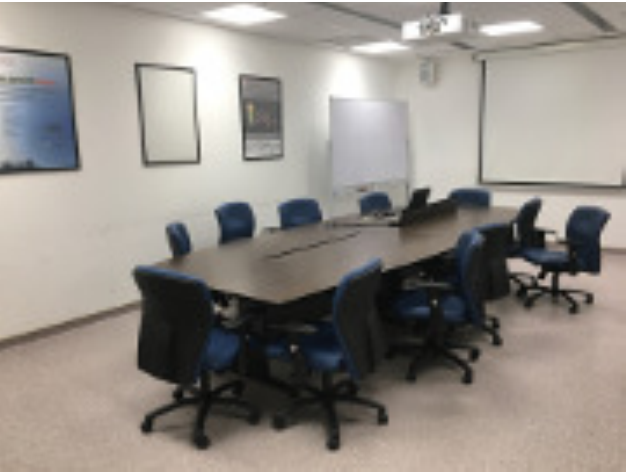}
  }
  \caption{ Examples of several room types.}
  \label{fig:diff_rooms}
\end{figure}

We conduct experiments in various types of rooms. Table~\ref{tab:description-rooms} summarizes all the residential, office, teaching, and museum rooms involved in our evaluation. We note that the design of the CNN hyperparameters, as presented in Section~\ref{subsec:hyper-settings}, is performed based on the data collected from the first 22 rooms summarized in Table~\ref{tab:description-rooms}, excluding the teaching rooms and museum halls. Fig.~\ref{fig:diff_rooms} shows the pictures of several types of rooms.

\begin{figure}
  \centering
  \subfigure[ No ambient music]
  {          
    \includegraphics[width=.44\textwidth]{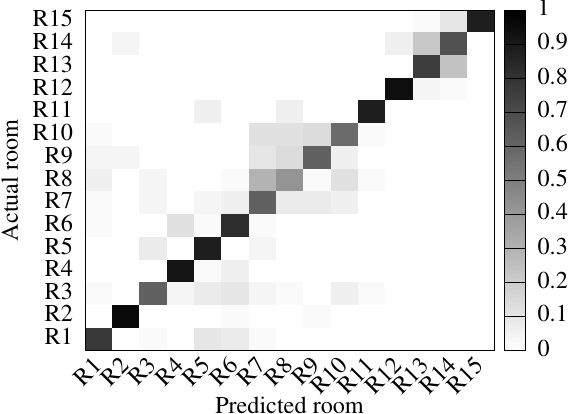}
    \label{fig:RS-mls-wo-music}
  }
  \hspace{.05\columnwidth}
  \subfigure[ With ambient music]
  {
    \includegraphics[width=.44\textwidth]{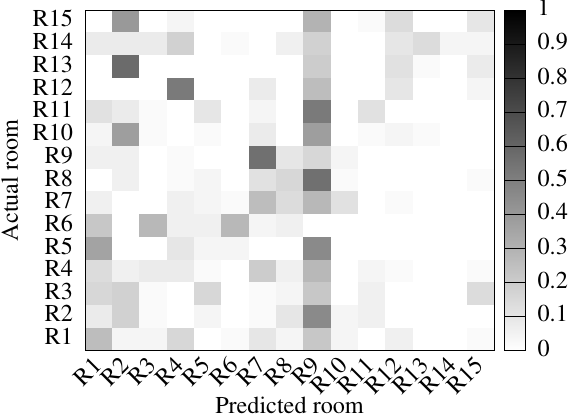}
    \label{fig:RS-mls-w-music}
  }
  \caption{ Confusion matrices of the original RoomSense.}
  \label{fig:roomsense-confusion}
\end{figure}

\begin{figure}
  \centering
  \subfigure[ No ambient music]
  {
    \includegraphics[width=.44\textwidth]{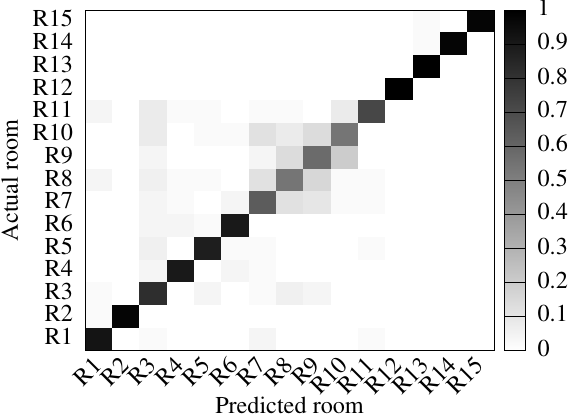}
    \label{fig:RS-st-wo-music}
  }
  \hspace{.05\columnwidth}
  \subfigure[ With ambient music]
  {          
    \includegraphics[width=.44\textwidth]{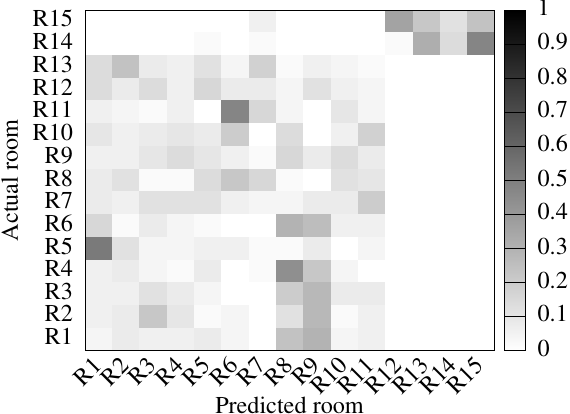}
    \label{fig:RS-st-w-music}
  }
  \caption{ Confusion matrices of single-tone broadband-MFCC RoomSense.}
  \label{fig:roomsense-st-bb-confusion}
\end{figure}

\begin{figure}
  \centering		
  \subfigure[ No ambient music]
  {          
    \includegraphics[width=.44\textwidth]{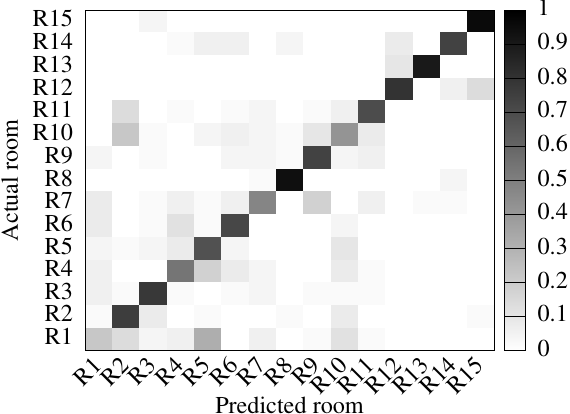}
    \label{fig:RS-st20-wo-music}
  }
  \hspace{.05\columnwidth}
  \subfigure[ With ambient music]
  {
    \includegraphics[width=.44\textwidth]{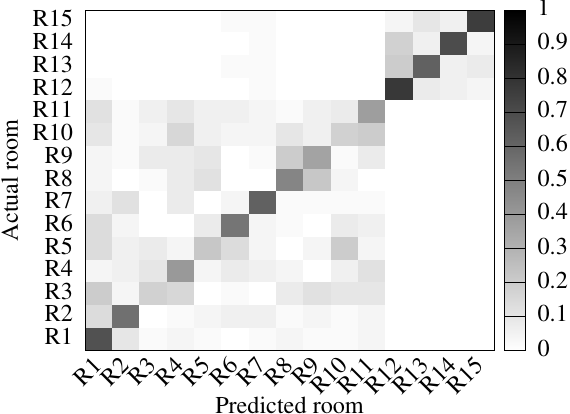}
    \label{fig:RS-st20-w-music}
  }
  \caption{ Confusion matrices of single-tone narrowband-MFCC RoomSense.}
  \label{fig:roomsense-st-nb-confusion}
\end{figure}

\begin{figure}
  \centering
  \subfigure[ No ambient music]
  {
    \includegraphics[width=.44\textwidth]{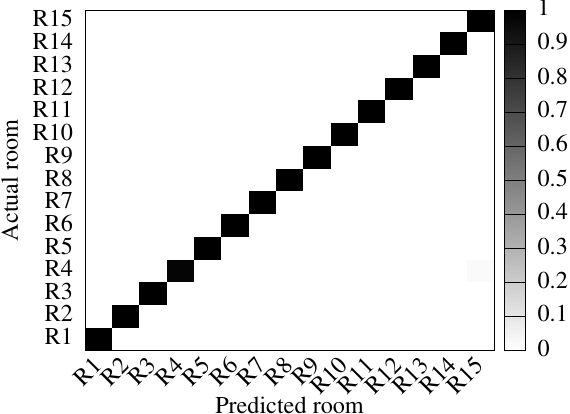}
    \label{fig:ours-wo-music}
  }
  \hspace{.05\columnwidth}
  \subfigure[ {\blue With ambient music }]
  {
    \includegraphics[width=.44\textwidth]{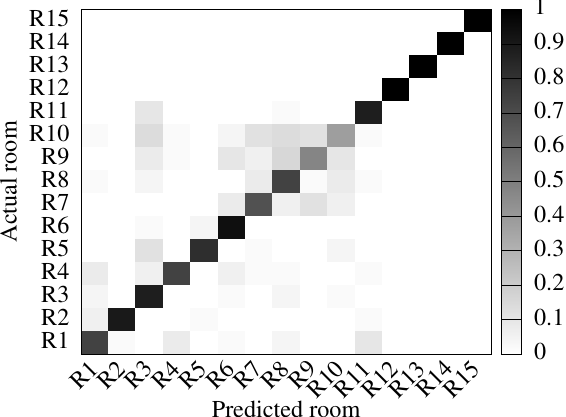}
    \label{fig:ours-w-music}
  }
  \caption{ Confusion matrices of RoomRecognize.}
  \label{fig:rr-confusion}
\end{figure}

\subsection{Evaluation Results in Residential, Office, and Teaching Rooms}

\subsubsection{Susceptibility to Interfering Sounds}
\label{subsubsec:interfering-sounds}

We evaluate the susceptibility of different room recognition approaches to interfering sounds. This set of experiments is conducted in 15 rooms chosen from the rooms listed in Table~\ref{tab:description-rooms}. {\blue We choose 4 bedrooms, 2 living rooms, 2 kitchens, 3 bathrooms, 2 visitor offices, and 2 meeting rooms. Note that in other rooms it is inconvenient or not allowed to play interfering sounds.}
In the experiment, we keep the rooms quiet when we collect training data for RoomRecognize, RoomSense, and its variants. When we test their recognition accuracy, we either keep the rooms quiet or play music using a laptop computer in the tested rooms.
Figs.~\ref{fig:RS-mls-wo-music} and \ref{fig:RS-mls-w-music} show the confusion matrices of the original RoomSense in the absence and presence of music, respectively.
The respective average recognition accuracy is 76\% and 39\%. Figs.~\ref{fig:RS-st-wo-music} and \ref{fig:RS-st-w-music} show the confusion matrices of the single-tone broadband-MFCC RoomSense in the absence and presence of music, respectively. The respective average recognition accuracy is 83\% and 27\%. Figs.~\ref{fig:RS-st20-wo-music} and \ref{fig:RS-st20-w-music} show the confusion matrices of the single-tone narrowband-MFCC RoomSense in the absence and presence of music, respectively. The respective average recognition accuracy is 69\% and 50\%.  Figs.~\ref{fig:ours-wo-music} and \ref{fig:ours-w-music} show the confusion matrices of RoomRecognize in the absence and presence of music, respectively. The respective average recognition accuracy is 100\% and 81\%. 

{\blue In Figs. ~\ref{fig:roomsense-confusion}, ~\ref{fig:roomsense-st-bb-confusion}, and ~\ref{fig:roomsense-st-nb-confusion}, the rooms R1-R4 are from a lab; the rooms R5-R10 and R11-R15 are from two different apartments. From Fig. ~\ref{fig:RS-mls-wo-music} and Fig. ~\ref{fig:RS-st-wo-music}, we can see some confusion blocks lumping together, e.g., the blocks representing R7-R10 and R13-R14 in Fig. ~\ref{fig:RS-mls-wo-music} and R7-R10 in Fig. ~\ref{fig:RS-st-wo-music}. This shows that the original RoomSense and the single-tone broadband-MFCC RoomSense make wrong classifications for the rooms from the same apartments. It suggests that these two approaches cannot well differentiate the rooms with similar floor and wall materials. In Fig. \ref{fig:RS-st20-wo-music} we can see that the confusion blocks are more dispersed, which means that the single-tone narrowband-MFCC RoomSense is better at recognizing rooms with similar furnishing materials. A possible reason for this improvement is that the narrowband-MFCC features carry less information about the tested room's furnishing material. When the ambient music is present, the confusion blocks appear randomly in Figs. ~\ref{fig:RS-mls-w-music}, ~\ref{fig:RS-st-w-music}, and ~\ref{fig:RS-st20-w-music}. This is because the ambient music is the main reason of the confusion.}

\begin{table} 
  \renewcommand{\arraystretch}{1.25}
  \caption{The average classification accuracy of different approaches in the absence and presence of music.}
  \label{tab:roomsense-vs-roomrecognize}
  \begin{tabular}{c|cc}
    \Xhline{1pt}
    Approach & In the absence of music & In the presence of music \\
    \hline
    Original RoomSense & 76\% & 39\% \\
    Single-tone broadband-MFCC RoomSense & 83\% & 27\% \\
    Single-tone narrowband-MFCC RoomSense & 69\% & 50\% \\
    RoomRecognize & 100\% & 81\% \\
    \Xhline{1pt}
  \end{tabular}
\end{table}

Table~\ref{tab:roomsense-vs-roomrecognize} summarizes the average recognition accuracy of different approaches in the absence and presence of music during the testing phase. From Table~\ref{tab:roomsense-vs-roomrecognize}, the original RoomSense and the single-tone broadband-MFCC RoomSense yield similar accuracy profiles. Note that both approaches use broadband-MFCC features. By narrowing the frequency band of the MFCC features to $[19.5, 20.5]\,\text{kHz}$, the single-tone narrowband-MFCC RoomSense achieves much better recognition accuracy in the presence of music. These comparisons show that using a narrow frequency band can significantly improve the system's robustness to interfering sounds. The single-tone narrowband-MFCC RoomSense performs worse than the other two RoomSense approaches in the absence of music. This is because the narrowband-MFCC features carry less information about the measured room than the broadband-MFCC features. From these results, we can see that the SVM used by RoomSense can hardly achieve a satisfactory Pareto frontier of recognition accuracy versus robustness against interfering sounds. Specifically, on one hand, due to the inferior learning capability, SVM needs echos in a broader frequency band to achieve a satisfactory recognition accuracy; on the other hand, the use of broadband audio will inevitably increase the system's susceptibility to interfering sounds. RoomRecognize has a 19\% accuracy drop in the presence of music, because the music may contain frequency components up to $20\,\text{kHz}$. However, compared with the SVM-based RoomSense, RoomRecognize gives a much improved Pareto frontier, owing to deep learning's strong ability to capture subtle differences in rooms' narrowband responses. We note that the 81\% average accuracy achieved by RoomRecognize in the presence of ambient music can be considered a worst-case result, because we introduce the interfering sounds in every room during the testing phase.




\subsubsection{Needed Training Data Volume}

Due to the deep models' massive parameters, a sufficient amount of training samples are critical to avoid overfitting. Fortunately, for the acoustics-based room recognition problem, the nearly automated training data collection process that repeatedly emits chirps and capture the room's echos can generate many training samples easily. This set of experiments evaluates the needed training data volume. Table~\ref{table:training-volumn-eval} shows the volume of training data for each room, the corresponding training data collection time, and the resulting accuracy in recognizing 22 rooms. We can see that, with 500 samples collected during 52 seconds in each room, the accuracy reaches 99.7\%.



\begin{table}
	\caption{Accuracy in recognizing 22 rooms under various training data volumes.}
	\label{table:training-volumn-eval}
	\begin{tabular}{lccccc}
		\toprule
		Volume of training data for each room (samples) & 100 & 250 & 375 & 437 & 500 \\
                \midrule

		Training data collection time for each room (seconds) & 10 & 26 & 39 & 45 & 52\\
		\midrule
		Room recognition accuracy (\%) & 83.15 & 90.5 & 91.3 & 94.5 & 99.7 \\
		\bottomrule
	\end{tabular}
\end{table}

\subsubsection{Impact of Phone Position and Orientation}

\begin{table}
  \begin{minipage}{.52\columnwidth}
    \centering
    \caption{Setting of data collection process for evaluating the impact of phone's position and orientation.}
    \label{table:evaluation-dataset}
    \begin{tabular}{cccc}
      \toprule
      Room & Size ($\text{m}^2$) & Number & Number of phone \\
      \multicolumn{2}{c}{} & of spots & orientations at each spot \\
      \midrule
      L1 & 10 & 6 & 6 \\
      L2 & 10 & 6 & 6 \\
      L3 & 12 & 8 & 6 \\
      L4 & 26 & 16 & 6 \\
      \bottomrule
    \end{tabular}
  \end{minipage}
  \hspace{.03\columnwidth}
  \begin{minipage}{.4\columnwidth}
    \centering
    \caption{Setting of the leave-one-spot-out cross validation. $x$-1 means that $x$ spots for training and one spot for testing.}
    \label{tab:position-per-room}
    \begin{tabular}{cc|ccccc}
      \Xhline{1pt}
      & & \multicolumn{5}{c}{\bf Spot density ($\text{m}^{-2}$)} \\
      & & $ \sfrac{1}{2} $ & $ \sfrac{1}{4} $ & $ \sfrac{1}{6} $ & $ \sfrac{1}{8} $ & $ \sfrac{1}{10} $ \\
      \hline
      \multirow{4}{*}{\rotatebox[origin=c]{90}{\bf Room}} & L1 & 5-1 & 3-1 & 2-1 & 1-1 & 1-1 \\
      & L2 & 5-1 & 3-1 & 2-1 & 1-1 & 1-1 \\
      & L3 & 6-1 & 3-1 & 2-1 & 2-1 & 1-1 \\
      & L4 & 13-1 & 7-1 & 4-1 & 3-1 & 3-1 \\
      \Xhline{1pt}
    \end{tabular}
  \end{minipage}
\end{table}

\begin{figure}
  \centering
  \begin{minipage}{.50\columnwidth}
    \includegraphics{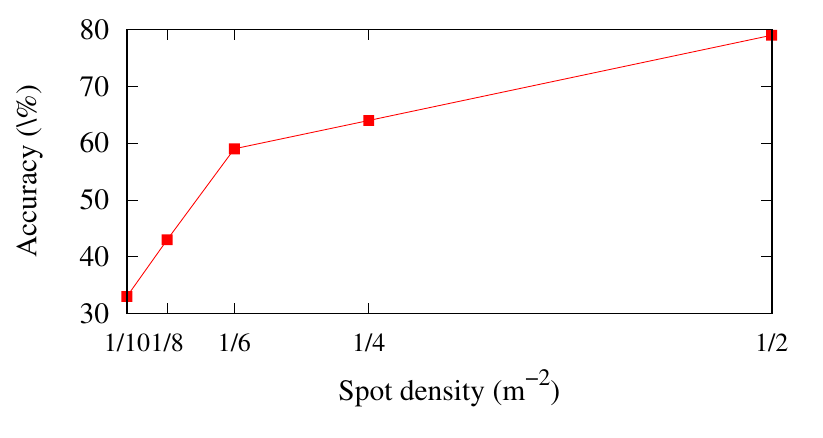}
    \vspace{-2em}
    \caption{\blue Leave-one-spot-out recognition accuracy vs. spot density for collecting training data.}
    \label{fig:leave-one-position-out-validation}
  \end{minipage}
  \hspace{.05\columnwidth}
  \begin{minipage}{.40\columnwidth}
    \vspace{-0.3in}
    \renewcommand{\arraystretch}{1.25}
    \centering
    \makeatletter\def\@captype{table}\makeatother
    \caption{Confusion matrix of leave-one-orientation-out cross validation. Average accuracy is 64\%.}
    \label{tab:leave-one-orientation-out-validation}
    \begin{tabular}{cc|cccc}
      \Xhline{1pt}
      & & \multicolumn{4}{c}{\bf Predicted room} \\
      & & L4 & L3 & L2 & L1 \\
      \hline
      \multirow{4}{*}{\rotatebox[origin=c]{90}{\bf Actual room}} & L4 & 0.65 & 0.14 & 0.11 & 0.10 \\
      & L3 & 0.15 & 0.60 & 0.10 & 0.15 \\
      & L2 & 0.03 & 0.14 & 0.61 & 0.22 \\
      & L1 & 0.04 & 0.03 & 0.23 & 0.70 \\
      \Xhline{1pt}
    \end{tabular}
  \end{minipage}
\end{figure}

We conduct a set of experiments in four rooms (L1 to L4 shown in Fig.~\ref{fig:lab-floor-plan}) to evaluate the impact of the phone's position and orientation on the performance of RoomRecognize. Specifically, we collect data using a phone at a total of 37 spots in the four rooms. The spots in a room are selected such that they are nearly evenly distributed in the room. At each spot, we collect 500 samples for each of six phone's orientations that are perpendicular to each other, i.e., front, back, left, right, up, and down.
Table ~\ref{table:evaluation-dataset} summarizes the settings of the data collection process.

In the first set of experiments, we conduct leave-one-spot-out cross validation to evaluate the impact of phone position on the performance of RoomRecognize. Specifically, out of a total of $n$ spots, we use the data collected at $n-1$ spots for training and the data collected at the remaining one spot for testing. Thus, the tested spot is not within the training data. We also vary $n$ to investigate the impact of the spot density for collecting training data (i.e., $1/(1-n)\,\text{m}^{-2}$) on the performance of RoomRecognize. Table~\ref{tab:position-per-room} summarizes the settings of the leave-one-spot-out cross validation experiments under different spot densities. In the leave-one-spot-out experiments, we consistently use the front orientation for the phone. Fig.~\ref{fig:leave-one-position-out-validation} shows the average leave-one-spot-out recognition accuracy versus the spot density for collecting training data. We can see that the recognition accuracy increases with the spot density. This means that, collecting training samples at more spots in each room will improve the performance of RoomRecognize, which is consistent with intuition. Two additional comments can be made regarding the results in Fig.~\ref{fig:leave-one-position-out-validation}. First, the leave-one-spot-out recognition accuracy is below 80\% when the spot density is up to 0.5 spot/$\text{m}^2$. As we select the spots evenly in each room, the leave-one-spot-out accuracy is the {\em worst-case} recognition accuracy (i.e., a lower bound) with respect to the impact of phone's position. Second, the number of spots in each room as summarized in Table~\ref{table:evaluation-dataset} achieve the spot density of 0.5 spot/$\text{m}^2$. Collecting data at several spots (e.g., 6 to 8 spots as in Table~\ref{table:evaluation-dataset}) in rooms with sizes of about $10\,\text{m}^2$ does not introduce significant overhead to the system trainer. While evenly selecting these spots is certainly preferred, at the end of this section, we will conduct another set of experiments in which the training data is collected when the phone carrier walks freely in each room.


Then, we evaluate the impact of phone orientation on RoomRecognize's performance by conducting a set of leave-one-orientation-out cross validation experiments. The data collected at all spots is used for training.
Table~\ref{tab:leave-one-orientation-out-validation} shows the confusion matrix of the leave-one-orientation-out cross validation. The average recognition accuracy is 64\%. This result shows that the phone orientation has larger impact on RoomRecognize, compared with the phone position. Note that the leave-one-out accuracy is the worst-case accuracy. By simply collecting training data for each of the six phone orientations, the worst case can be avoided.

In the third set of experiments, we assess RoomRecognize's performance when the phone carrier walks freely in the measured rooms and rotate the phone randomly during walking. Out of totally 1,000 samples collected during a two-minute free walk in each room, 500, 250, and 250 samples are used for training, validation, and testing, respectively. The average room recognition accuracy is 87\%. This result shows that RoomRecognize can achieve a satisfactory recognition accuracy without imposing strict requirements to the training data collection process. The system trainer can follow a general guideline of walking in the walkable areas of a room and rotating the phone randomly during the training data collection process.


\subsubsection{Scalability}

\begin{figure}
  \begin{minipage}[t]{.4\textwidth}
    \includegraphics{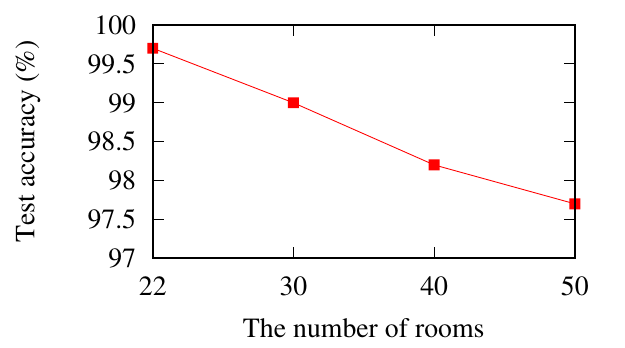}
    \caption{\blue Test accuracy vs. number of rooms.}
    \label{fig:num-classes-vs-accuracy}
  \end{minipage}
  \hfill
  \begin{minipage}[t]{.55\textwidth}
    \centering
    \includegraphics{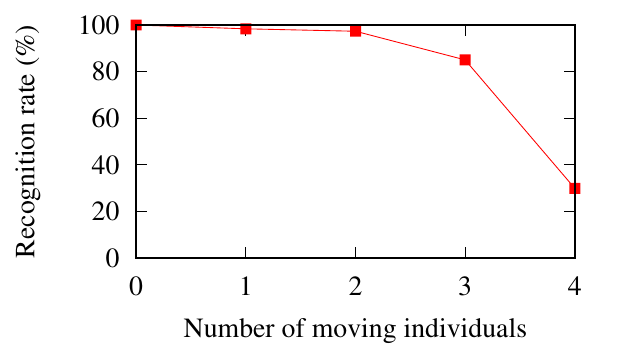}
    \caption{\blue Recognition rate vs. the number of moving individuals.}
    \label{fig:acc_vs_people}
  \end{minipage}
\end{figure}

In Section~\ref{subsec:hyper-settings}, we designed CNN-C based on data collected from 22 rooms. In this section, we evaluate how well CNN-C can scale with the number of rooms (i.e., the number of classes). We collect data from additional 28 rooms. Thus, we have data from a total of 50 rooms. Among 1,000 samples collected from each room, 500, 250, and 250 samples are included into the training, validation, and testing data sets, respectively. Fig.~\ref{fig:num-classes-vs-accuracy} shows the average test accuracy versus the number of rooms. We can see that the test accuracy decreases with the number of rooms, which is consistent with intuition. However, RoomRecognize still gives a 97.7\% accuracy in classifying 50 rooms. The scale of 50 rooms is satisfactory for a range of application scenarios, i.e., recognizing museum exhibition chambers, wards of a hospital department, etc. 

{ \subsubsection{Impact of Surrounding Moving People}
  \label{subsubsec:moving-people}

  We conduct an experiment in L3 shown in Fig.~\ref{fig:lab-floor-plan} to evaluate the impact of surrounding moving people on the performance of RoomRecognize. As described in Table ~\ref{tab:description-rooms}, L3 has a size of about $7\,\text{m}^2$. Training and testing data sets are collected from the same spot in the room. When collecting the training data, we keep the room empty and quiet. The training data for this room and 21 other rooms are used to train RoomRecognize. In the testing phases, we invite a number of volunteers to walk randomly in the room.
  Fig.~\ref{fig:acc_vs_people} shows the probability that the L3 is correctly recognized versus the number of moving individuals in the room. We can see that the recognition rate decreases with the number of surrounding moving individuals. This is because an individual person can reflect the inaudible acoustic signal emitted by a RoomRecognize phone. Thus, if the people move in a room, the temporal process of the room's response to the chirp will change and RoomRecognize's performance will drop. Intuitively, the degree of the change and RoomRecognize's performance degradation increase with the number of moving individuals. The result also shows that in a room with a size of $7\,\text{m}^2$, two moving individuals result in a recognition rate drop of 2.7\% only. In Section~\ref{subsec:museum-eval}, the impact of moving people on RoomRecognize will be also evaluated in a museum.
}

{ \subsubsection{Impact of Changes of Movable Furniture}
  \label{subsubsec:movable-furniture}
  We conduct an experiment in L3 to evaluate the impact of changes of movable furniture on RoomRecognize's performance. In this room, several chairs and a round table are movable furniture. Fig.~\ref{fig:l3-original} shows the original furniture layout. The training data for this room collected under the setting shown in Fig.~\ref{fig:l3-original} and the data collected from other 21 rooms are used to train RoomRecognize. During the testing phases, we move the chairs and the table around in the room, remove most chairs, and add more chairs, as shown in Figs.~\ref{fig:l3-move}, \ref{fig:l3-remove}, \ref{fig:l3-add}, respectively. For the settings in Figs.~\ref{fig:l3-move}, \ref{fig:l3-remove}, \ref{fig:l3-add}, the probabilities of correctly recognizing L3 are 100\%, 92\%, and 100\%, respectively. This experiment shows that the changes of movable furniture in a room may affect the performance of RoomRecognize, since the furniture also affects the reverberation process of the inaudible sounds. However, the changes of the movable furniture, as shown in Fig.~\ref{fig:changes_in_L3}, do not subvert RoomRecognize.

  As permanent facilities in a room are often bulky, their changes may affect the performance of RoomRecognize significantly. Thus, RoomRecognize should be retrained if any new permanent facility is added and/or any existing permanent facility is changed/removed.
}

\begin{figure}
  \subfigure[ Original layout.]{
    \includegraphics[width=.23\textwidth]{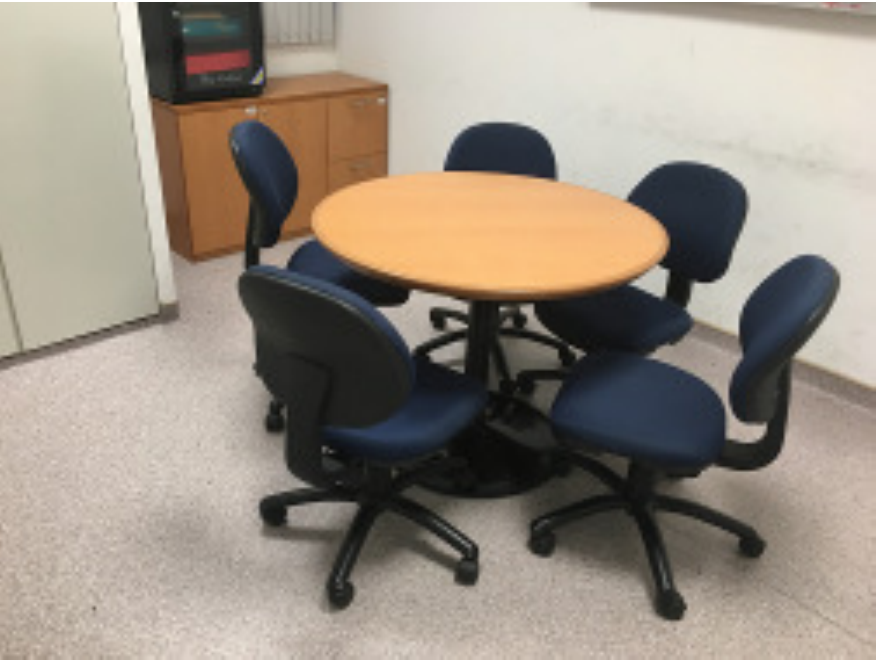}
    \label{fig:l3-original}
  }
  \subfigure[ Chairs and table moved.]{
    \includegraphics[width=.23\textwidth]{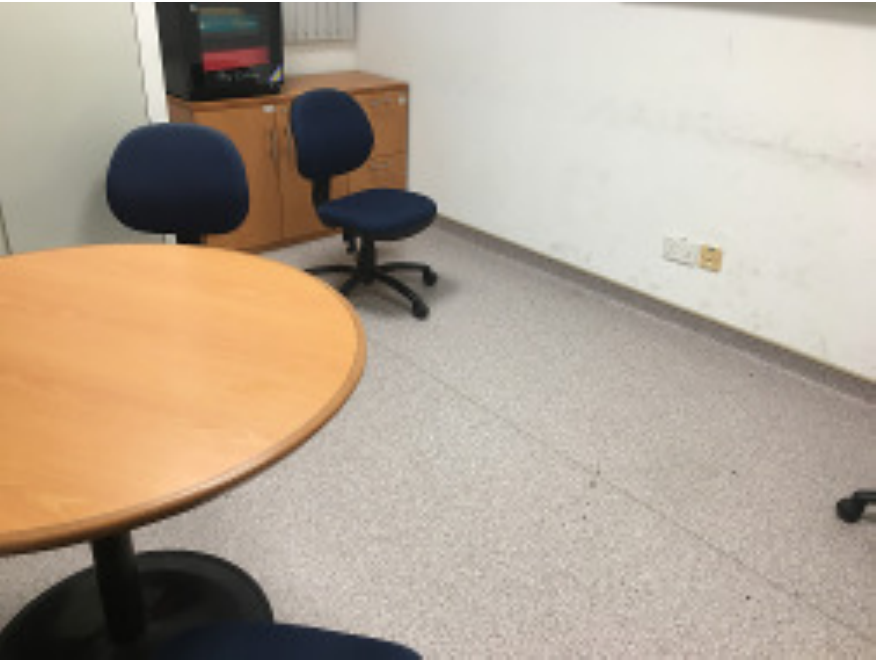}
    \label{fig:l3-move}
  }
  \subfigure[ Chairs removed.]{
    \includegraphics[width=.23\textwidth]{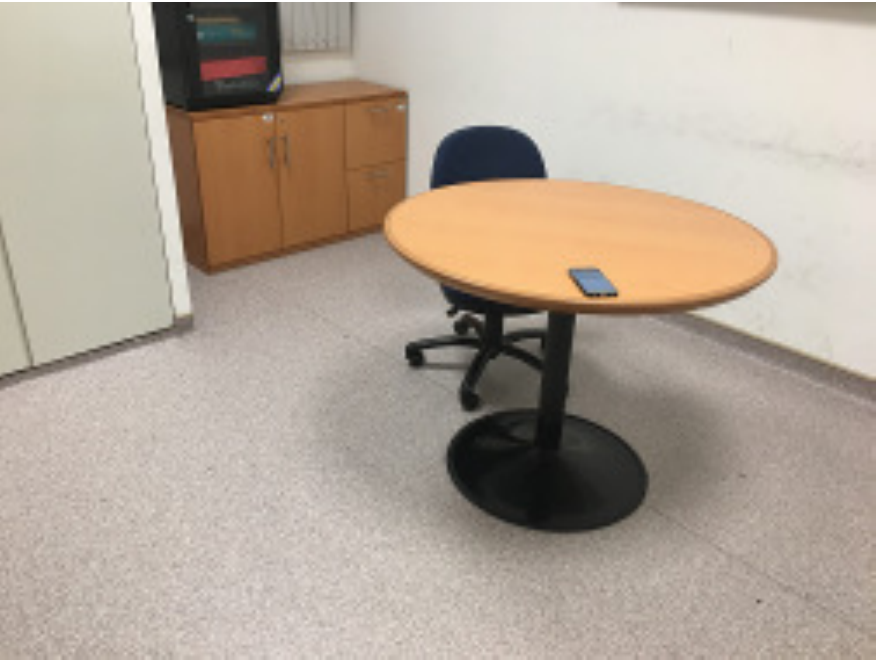}
    \label{fig:l3-remove}
  }
  \subfigure[ More chairs added.]{
    \includegraphics[width=.23\textwidth]{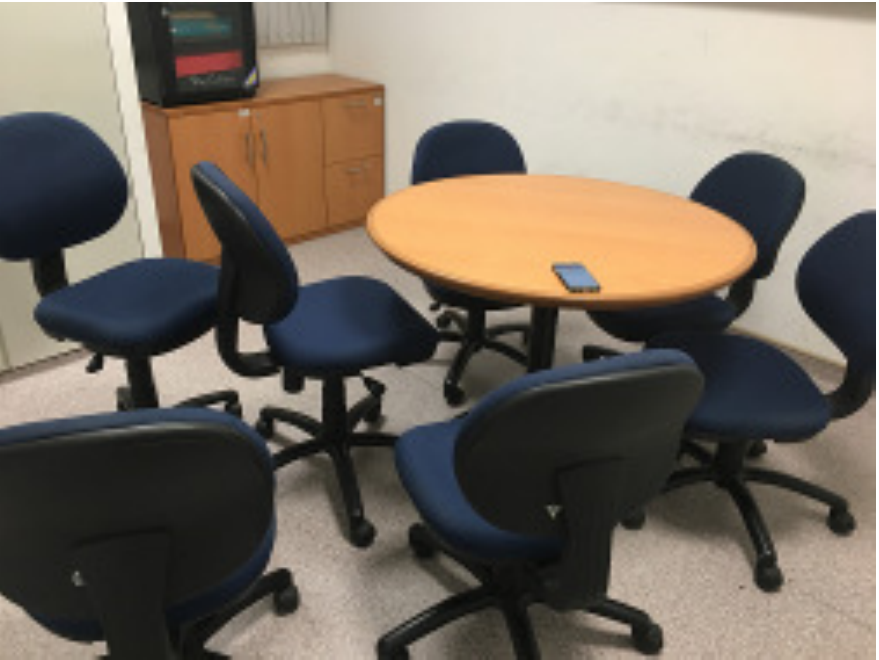}
    \label{fig:l3-add}
  }
  \caption{ Furniture changes in L3.}
  \label{fig:changes_in_L3}
\end{figure}

{
  \subsubsection{Evaluation in Similar Rooms}
  \label{subsubsec:similar-rooms}

  \begin{figure}
  \subfigure[ Teaching room 1]
  {
    \includegraphics[width=0.18\textwidth]{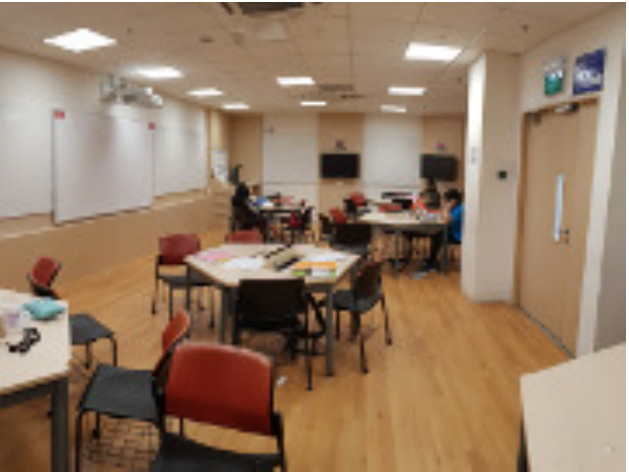}
  }
  \subfigure[ Teaching room 2]
  {
    \includegraphics[width=0.18\textwidth]{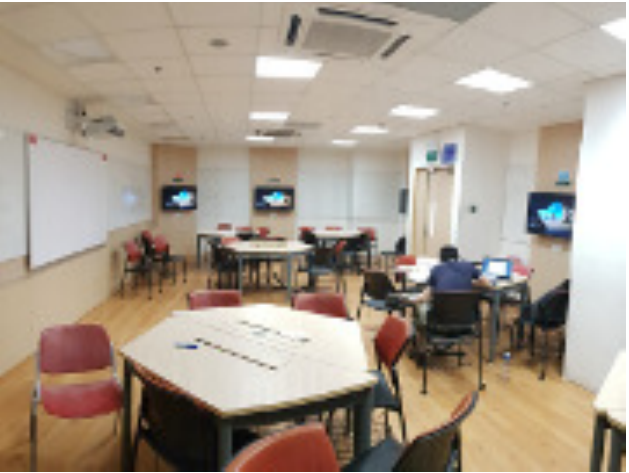}
  }
  \subfigure[ Teaching room 3]
  {
    \includegraphics[width=0.18\textwidth]{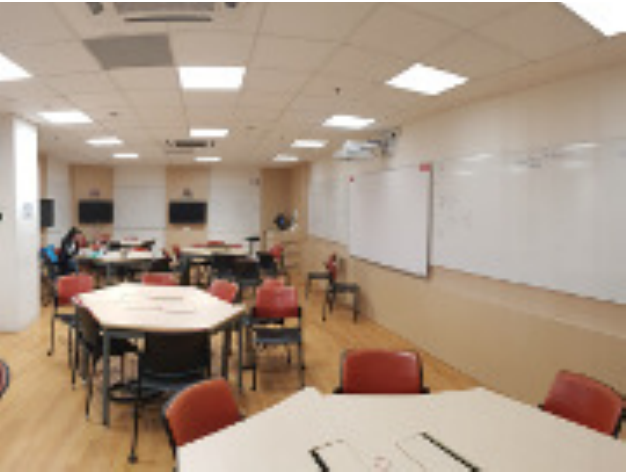}
  }
  \subfigure[ Teaching room 4]
  {
    \includegraphics[width=0.18\textwidth]{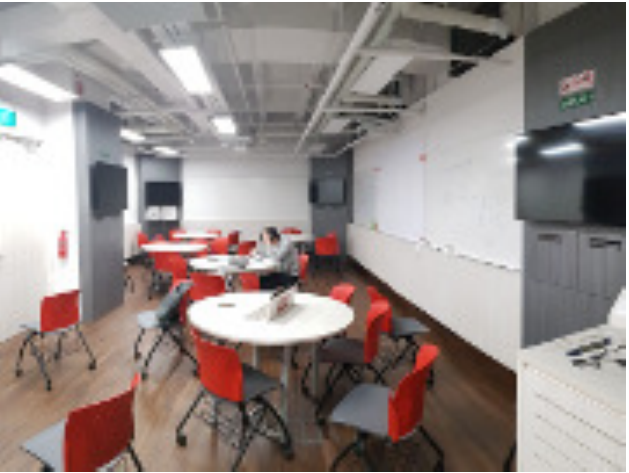}
  }
  \subfigure[ Teaching room 5]
  {
    \includegraphics[width=0.18\textwidth]{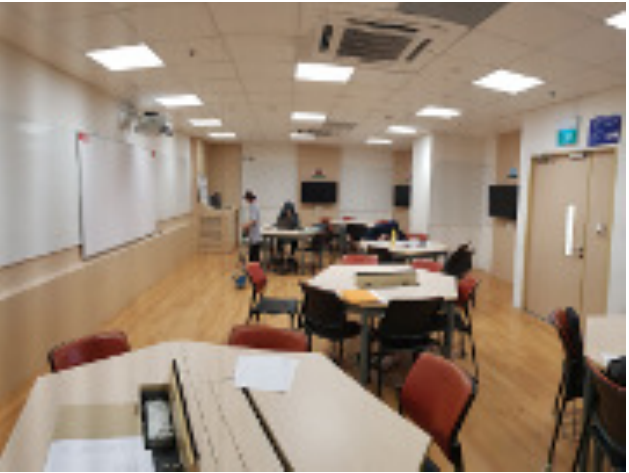}
  }
  \caption{ Examples of similar rooms.}
  \label{fig:similar_rooms}
\end{figure}

We evaluate the performance of RoomRecognize in recognizing similar rooms. We select 10 teaching rooms in a university that have similar sizes, layouts, furniture, and furnishing. Fig.~\ref{fig:similar_rooms} shows the pictures taken from five of them.
In each room, we select one spot to collect 500, 250, 250 samples as training, validation, and testing data. Fig.~\ref{fig:RR_similar_rooms} shows the confusion matrix of RoomRecognize in recognizing the 10 similar rooms. From the confusion matrix, each teaching room receives similar recognition rate. RoomRecognize achieves an average accuracy of 88.9\%. Compared with RoomRecognize's nearly perfect accuracy obtained in Section~\ref{subsubsec:interfering-sounds}, the high similarity of the rooms in this experiment results in an accuracy drop. However, RoomRecognize performs better than the single-tone narrowband-MFCC RoomSense in recognizing these 10 similar rooms. Specifically, Fig.~\ref{fig:singletone-nb-MFCC-RS_similar_rooms} shows the confusion matrix of the single-tone narrowband-MFCC RoomSense. RoomSense achieves an average accuracy of 72\% only. The better performance of RoomRecognize is due to deep learning's better capability in capturing subtle differences in rooms' narrowband responses.
}

\begin{figure}
  \centering
  \subfigure[ RoomRecognize]
  {
    \includegraphics[width=.44\textwidth]{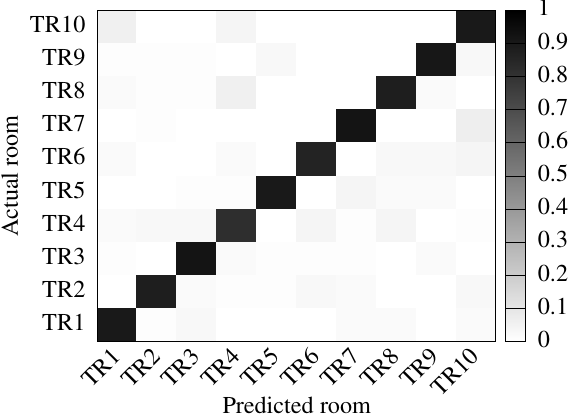}
    \label{fig:RR_similar_rooms}
  }
  \hspace{.05\columnwidth}
  \subfigure[ Single-tone narrowband-MFCC RoomSense]
  {
    \includegraphics[width=.44\textwidth]{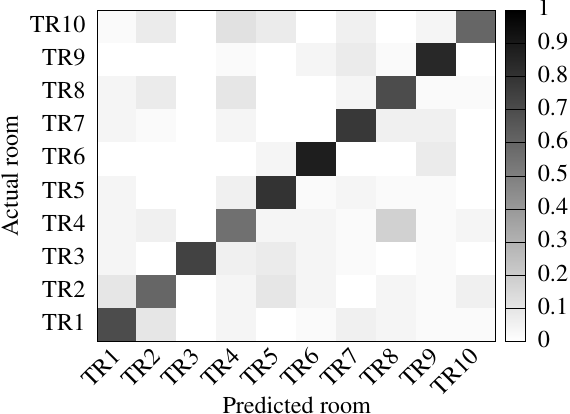}
    \label{fig:singletone-nb-MFCC-RS_similar_rooms}
  }
  \caption{ Confusion matrices of RoomRecognize and single-tone narrowband-MFCC RoomSense in recognizing 10 similar teaching rooms (TR).}
\end{figure}

{\blue \subsubsection{Run-Time Room Recognition Latency}
\label{subsubsec:runtime-latency}
We conduct a set of experiments to evaluate the run-time latency for executing the CNNs. Table~\ref{table:latency-cnns} summarizes the run-time latency of various CNNs listed in Table~\ref{table:DRR-configuration}. The latencies are for processing a single sample. The results show that the run-time latency for processing a sample is a few microseconds only. Note that these CNNs are trained using the data collected from the first 22 rooms listed in Table~\ref{tab:description-rooms}. We also evaluate how the run-time recognition latency scales with the number of rooms that the CNN is trained to handle. First, we train the CNN-B, -C, and -D from Table~\ref{table:DRR-configuration} with different numbers of rooms and achieve roughly the same recognition accuracy (97\%). Table~\ref{table:latency-vs-classes} shows the run-time latency of these CNNs in processing a sample. We can see that the CNN-C that can recognize 50 rooms takes doubled time to process a sample than CNN-B and CNN-D that can recognize 10 and 11 rooms. Second, we train CNN-C to recognize 10 to 50 rooms. Fig.~\ref{fig:latency-no_of_rooms} shows the run-time latency of CNN-C in processing a sample when it is trained to handle different numbers of rooms. We can see that the run-time latency increases with the number of rooms. In particular, when more than 20 rooms are considered, the run-time latency exhibits a linear relationship with the number of rooms. Nevertheless, as the CNN's run-time latency is at the microseconds level, the end-to-end latency experienced by the user will be dominated by the network communication delays that are generally tens of milliseconds. The above results show that our approach is efficient at run time.

\begin{table}
  \centering
  \caption{\blue Run-time recognition latency of various CNNs listed in Table 2.}
  \label{table:latency-cnns}
  \begin{tabular}{lccccccc}
    \toprule
    CNN- & A & B & C & D & E & F & G\\
    \midrule
    Run-time latency ($\mu$s) & 6.45 & 3.23 & 7.04 & 7.04 & 6.16 & 3.23 & 3.81\\
    \bottomrule
  \end{tabular}
\end{table}

\begin{table}
  \centering
  \caption{\blue Run-time latency of CNNs with same recognition accuracy.}
  \label{table:latency-vs-classes}
  \begin{tabular}{lccc}
    \toprule
    CNN- & B & C & D \\
    \midrule
    Number of rooms & 10 & 50 & 11 \\
    Run-time latency ($\mu$s) & 3.52 & 7.62 & 3.23 \\
    \bottomrule
  \end{tabular}
\end{table}

\begin{figure}
  \centering
  \includegraphics[width=.5\textwidth]{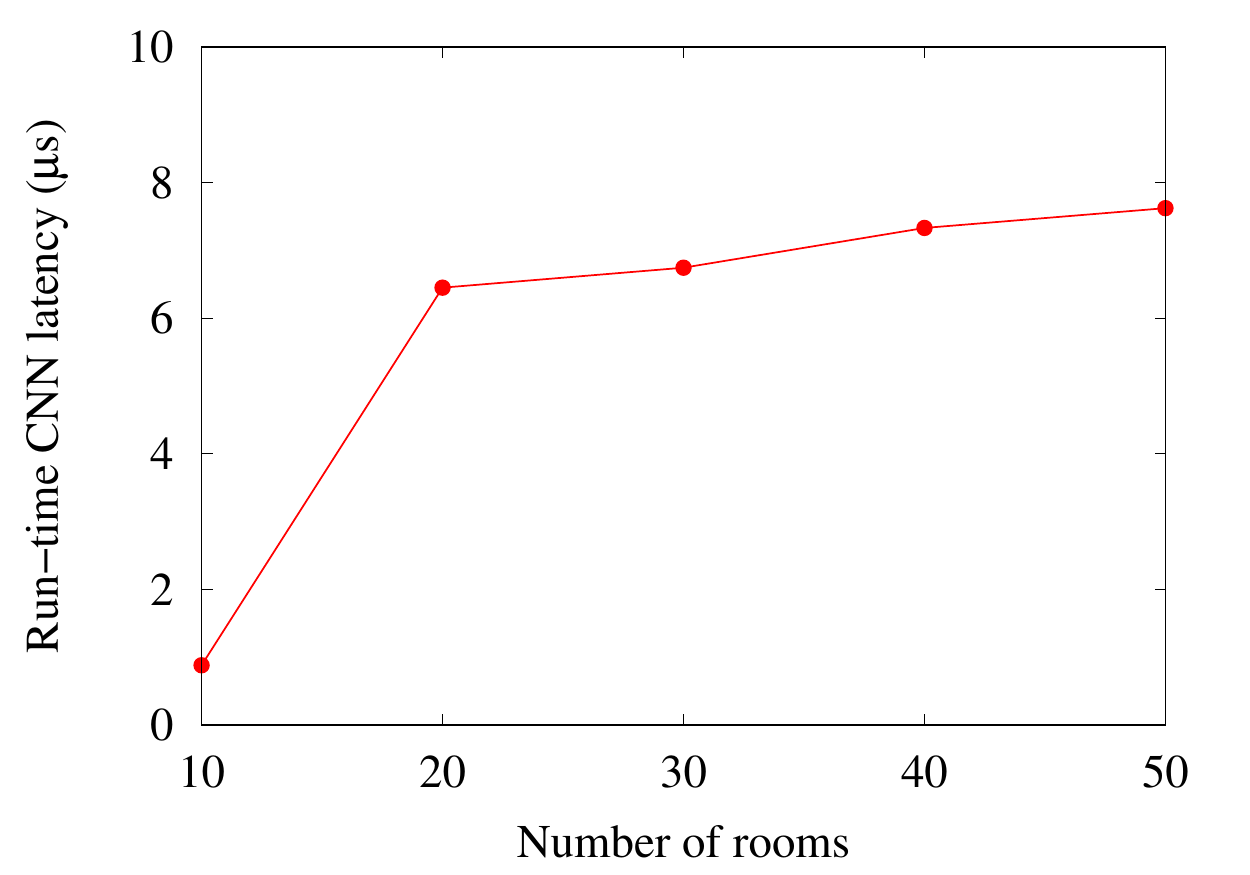}
  \caption{\blue Room recognition latency vs. the number of rooms.}
  \label{fig:latency-no_of_rooms}
\end{figure}

}

\subsection{Evaluation Results in Two Museums}
\label{subsec:museum-eval}

\begin{figure}
  \begin{minipage}{.6\columnwidth}
    \includegraphics[width=\textwidth]{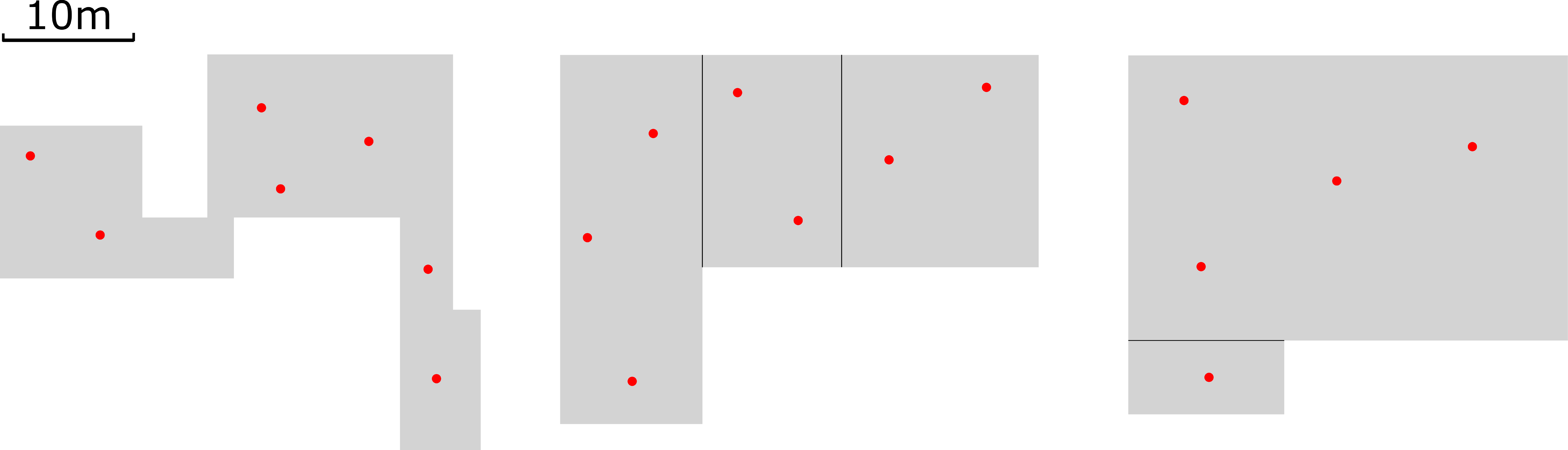}
    \caption{Museum-A floor plan and data collection spots (red points).}
    \label{fig:NUSM_floor_plan}
  \end{minipage}
  \hspace{.1\columnwidth}
  \begin{minipage}{.28\columnwidth}
    \includegraphics[width=\textwidth]{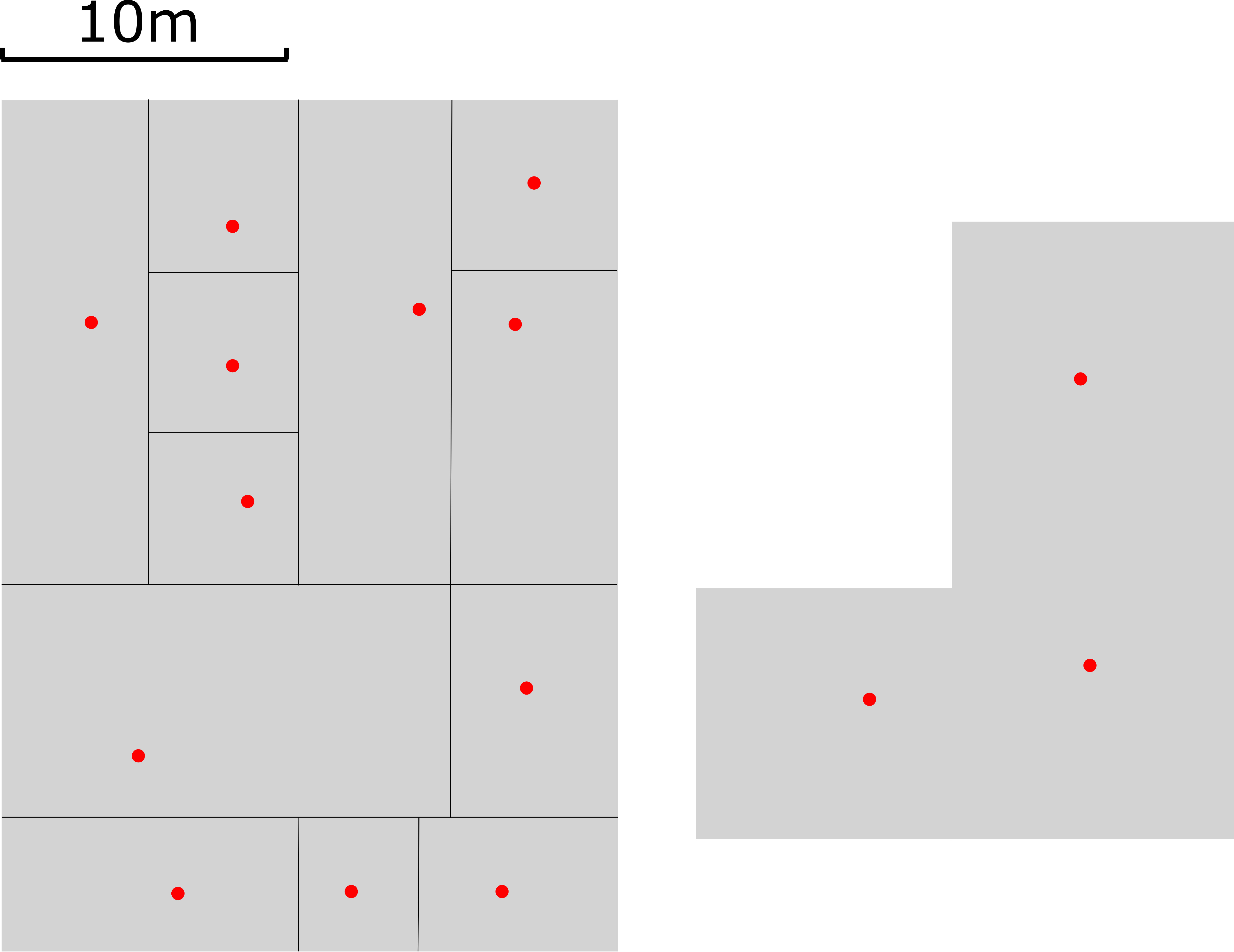}
    \caption{Museum-B floor plan and data collection spots (red points).}
    \label{fig:LKCNHM_floor_plan}
  \end{minipage}
\end{figure}
      
We also evaluate RoomRecognize in two museums, Museum-A and Museum-B.
Figs.~\ref{fig:NUSM_floor_plan} and \ref{fig:LKCNHM_floor_plan} show the floor plans of the two museums. Note that RoomRecognize can be naturally applied to the museums that consist of many small exhibition chambers. Differently, Museum-A and Museum-B consist of large exhibition halls. While we can apply RoomRecognize to recognize different halls, we are interested in investigating whether RoomRecognize can recognize spots in the large exhibition halls.

In front of each exhibition item, we select a spot to collect data samples. When collecting data at a spot, the phone carrier holds the phone in one hand, and rotates randomly the phone to multiple orientations, and walks around the spot without exceeding a distance of about one meter. Out of 1,000 samples collected at each spot during a two-minute process, 500, 250, and 250 samples are used as training, validation, and testing data. In total, 19 spots and 15 spots are selected from three and two exhibition halls in Museum-A and Museum-B, respectively. The locations of these spots are illustrated in Figs.~\ref{fig:NUSM_floor_plan} and \ref{fig:LKCNHM_floor_plan}. The average spot recognition accuracy in Museum-A and Museum-B is 99\% and 89\%, respectively. During the data collection process in Museum-A, there was a limited number of visitors walking around. In contrast, during the data collection process in Museum-B, there was background music and the museum was crowded. Thus, we believe that RoomRecognize's performance drop in Museum-B is caused by the moving crowd, as explained in Section~\ref{subsubsec:moving-people}. Nevertheless, RoomRecognize still gives a satisfactory accuracy of 89\%.




\section{Discussions}
\label{sec:discuss}


From the experiment results in Section~\ref{subsubsec:moving-people} and Section~\ref{subsec:museum-eval}, moving people in the target rooms results in RoomRecognize's performance drop. This is because of the susceptibility of the acoustic signals to the barriers such as human bodies. To address this issue, other sensing modalities that are not affected by nearby human bodies can be incorporated in RoomRecognize. One promising sensing modality is geomagnetism. The fusion of multi-modal sensing data for room recognition needs further study. One possible fusion method is to yield the most confident classification result that is made based on a single sensing modality. In our future work, we will also investigate whether a unified deep learning model that takes both inaudible echo data and geomagnetism data as inputs can improve the robustness of RoomRecognize against nearby moving people.

As shown in Section~\ref{subsubsec:similar-rooms}, recognizing a considerable number of similar rooms is worth further study, although RoomRecognize has outperformed the state-of-the-art approach in recognizing similar rooms. A possible approach to further improve RoomRecognize's performance is to use multiple inaudible tones within the phone audio system's capability, e.g., from $20.0\,\text{kHz}$ to $20.6\,\text{kHz}$ for Samsung Galaxy S7 as shown in Fig.~\ref{fig:inaudibles-power}. The room's responses at different frequencies will increase the amount of information about the room, therefore potentially improving RoomRecognize's performance in discriminating similar rooms.

\section{Conclusion}
\label{sec:conclude}

This paper presented the design of a room-level indoor localization approach based on the measured room's echos in response to a two-millisecond single-tone inaudible chirp emitted by a smartphone's loudspeaker. Our approach records audio in a narrow inaudible band for 0.1 seconds only to preserve the user's privacy. To address the challenges of limited information carried by the room's response in such a narrow band and a short time, we applied deep learning to effectively capture the subtle differences in rooms' responses. Our extensive experiments based on real echo data traces showed that a two-layer CNN fed with the spectrogram of the echo achieves the best room recognition accuracy. Based on the CNN, we designed {\em RoomRecognize}, a cloud service and its mobile client library to facilitate the development of mobile applications that need room-level localization. Extensive evaluation shows that RoomRecognize achieves accuracy of 99.7\%, 97.7\%, 99\%, and 89\% in differentiating 22 and 50 residential/office rooms, 19 spots in a quiet museum, and 15 spots in a crowded museum, respectively. Moreover, compared with Batphone \cite{tarzia2011indoor} and RoomSense \cite{rossi2013roomsense}, two acoustics-based room recognition systems, our CNN-based approach significantly improves the Pareto frontier of recognition accuracy versus robustness against interfering sounds.


\begin{acks}

The work is supported in part by an NTU Start-up Grant and an NTU CoE Seed Grant. We acknowledge the support of NVIDIA Corporation with the donation of the Quadro P5000 GPU used in this research.

\end{acks}

\bibliographystyle{ACM-Reference-Format}
\bibliography{references}

\end{document}